\documentclass[manuscript]{acmart}
\usepackage{tabularx}
\usepackage{array}
\usepackage{booktabs}
\setcopyright{none}
\hypersetup{pdfauthor={Yuhang Jin, Mengfei Chen, Hanwei Zheng, Songruoyao Wu, Kejun Zhang},pdftitle={Creative Intent as an Evolving Shared State: Supporting Cross-Stage Human–AI Video Co-Creation}}
\title{Creative Intent as an Evolving Shared State: Supporting Cross-Stage Human–AI Video Co-Creation}
\author{Yuhang Jin}
\author{Mengfei Chen}
\author{Hanwei Zheng}
\author{Songruoyao Wu}
\author{Kejun Zhang}
\renewcommand{\shortauthors}{Jin et al.}

\begin{document}
\begin{abstract}
As generative AI becomes increasingly involved in video creation, creators face challenges in expanding, translating, and negotiating creative intent across stages. We therefore view creative intent as an evolving shared state throughout human–AI co-creation. A multi-stage design was employed: semi-structured interviews with 12 creators informed 3 design goals, followed by expert co-design and the development of VideoMaker. VideoMaker organizes creation into a six-stage workflow that progressively translates creative intent. At key stages, creators negotiate with AI through the Seed Panel, explicitly confirming, revising, and updating intent. A shared AI Agent, supported by the Multi-style Script Library and Cinematic Language Knowledge Base, enables differentiated expansion of initial intent and faithful translation. A within-subjects study with 30 participants showed that VideoMaker effectively supported the expansion and cross-stage translation of creative intent, with stronger intent preservation, while negotiated boundaries between creator-confirmed intent and AI initiative remained difficult to sustain.
\end{abstract}


\maketitle

\section{Introduction}
\label{sec:1}

With the rapid development of generative AI \cite{gupta2024photorealistic,cao2025aigc,he2025historical}, its role in video has expanded from media synthesis to content understanding, creative planning, video production, editing, and distribution, increasingly spanning the entire creative process \cite{zhou2024surveyvideo,xia2022synthesis}. Across filmmaking, short-form video, music video, educational video, and news video, AI has been applied to shot organization, scriptwriting, media generation, and post-production \cite{wang2024filmmaking,kim2024unlocking,anderson2025making,lee2025mvprompt,baek2026pedacogen,wang2024reelframer}. Accordingly, creator–AI interaction is shifting from isolated tool use toward sustained collaboration throughout the creative process.\par

AI-assisted video creation is a multi-stage, multi-representational, nonlinear, and iterative process \cite{lee2025mvprompt,yeh2026vidmento,li2026creatorcentric,hu2026previzwhiz,wang2026script2screen,li2026sketchdynamics}. Creators first develop vague ideas into clearer creative directions, progressively translate them into scripts and storyboards, refine them through audiovisual previsualization and generated results, and iteratively optimize the work \cite{antony2024id8,liu2026textnative,hu2026previzwhiz,wang2026script2screen,li2026sketchdynamics}. However, deeper AI involvement introduces new challenges. AI generation is shaped by existing data patterns, which may lead to homogenization of AI-generated content and constrain the differentiated expansion of creators’ initial intent \cite{wadinambiarachchi2024effects,doshi2024generative,kim2024unlocking}. Turning an initial idea into video also requires repeated translations across representations, during which AI may reinterpret the original intent when supplementing and specifying information \cite{shin2024lostmelody,liu2026textnative,li2026creatorcentric,wang2026script2screen,hu2026previzwhiz}. Meanwhile, excessive human specification can restrict AI’s generative space, whereas excessive AI autonomy may weaken creators’ authority over key creative decisions. Rather than treating creative intent as a fixed input to be preserved, we view it as an evolving state that must be continuously expanded, translated, and negotiated throughout human–AI co-creation.\par

Prior work has addressed these challenges from technical \cite{zhou2024surveyvideo,ma2025controllable,guo2024animatediff,jiang2025vace,ma2025videodiffusion}, narrative \cite{zheng2024videogen}, and practical \cite{pan2025creative,kim2024unlocking} perspectives. From a human–AI co-creation perspective, existing tools mainly adopt four approaches. Multi-alternative exploration and comparison support divergent exploration \cite{huh2025videodiff,liu2026textnative,antony2024id8,yeh2026vidmento}; structured creative scaffolds and intermediate representations, such as storylines, scripts, and storyboards, progressively specify creative intent \cite{antony2024id8,lee2025mvprompt,wang2024reelframer,baek2026pedacogen,cao2025compositional}; multimodal and novel interactions—including VR, sketches, 3D previsualization, and text–video mappings—support the expression of complex intent \cite{wen2025seewhat,hu2026previzwhiz,li2026sketchdynamics,tilekbay2024expressedit,wang2026script2screen,wang2026rewriting,lin2026visuallyrics,liu2026textnative}; suggestive assistance and creator-led decision making help maintain creator control over key decisions \cite{wang2024podreels,wang2024lave,wang2024reelframer,yeh2026vidmento,baek2026pedacogen,savytski2026bridging}. Despite these advances, most approaches focus on individual stages or local tasks, with limited consideration of how creative intent can function as a shared state across the entire video creation process.\par

To further understand how these challenges arise in real creative practice, we conducted a formative study. Semi-structured interviews with 12 creators revealed a central challenge: how to construct creative intent as an evolving shared state. This challenge manifested in three dimensions—intent expansion, translation, and negotiation—and informed three design goals: supporting differentiated expansion around creative intent (DG1); supporting the progressive and faithful translation of creative intent across stages and representations (DG2); and supporting continuously negotiable boundaries between creator-confirmed intent and AI initiative (DG3). We then invited three experts to a co-design process to translate these goals into concrete mechanisms. This process led to the development of a Multi-style Script Library for DG1, a six-stage AI Video Authoring Workflow and Cinematic Language Knowledge Base for DG2, and the Seed Panel for DG3, informed by prior theory \cite{clark1991grounding,scaife1996external,horvitz1999mixedinitiative,boden2004creative,fauconnier2002way}.\par

\begin{figure}[!htbp]
\centering
\includegraphics[width=\linewidth]{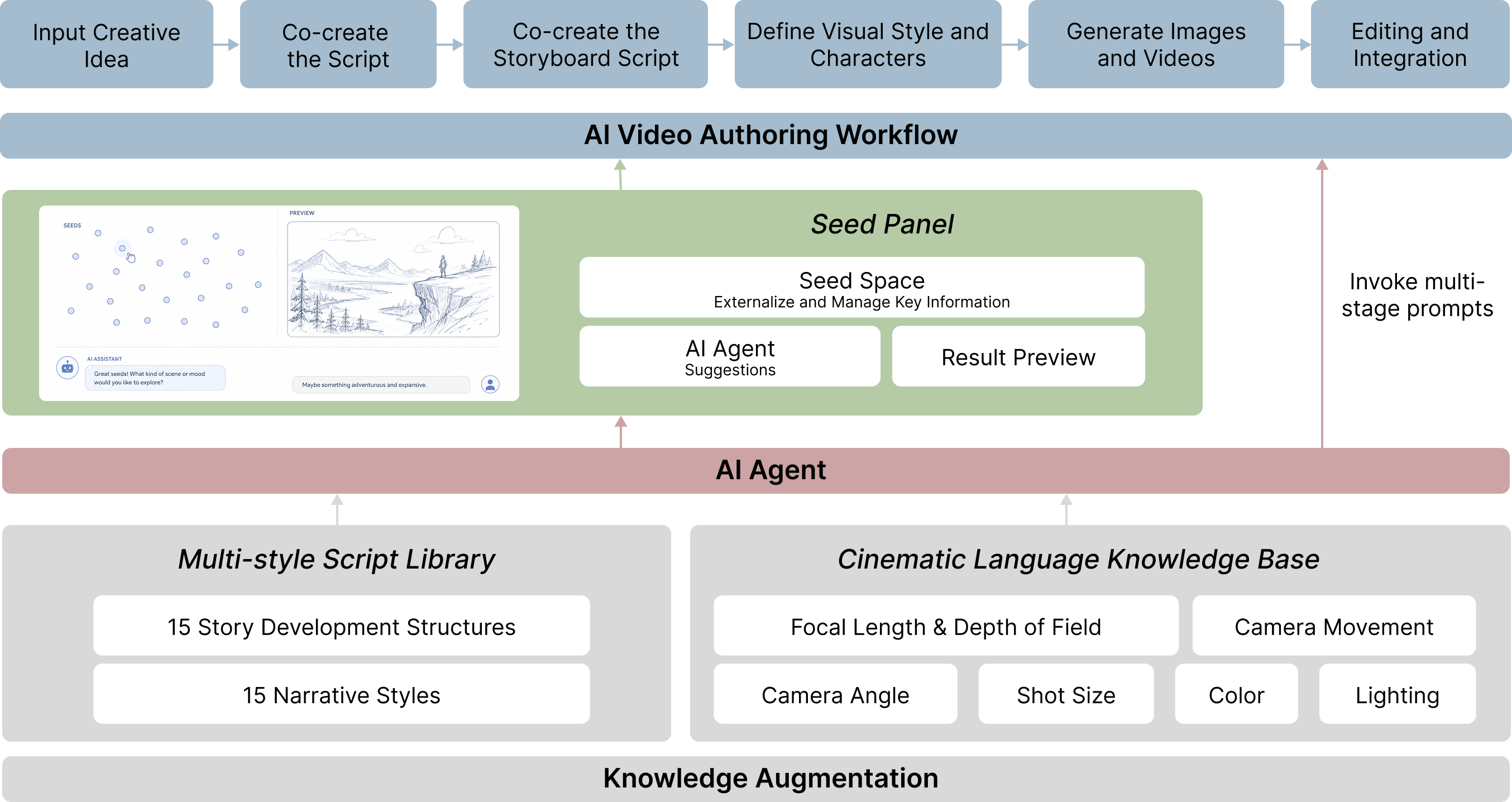}
\caption{Overview of VideoMaker. The system operationalizes creative intent as an evolving shared state through a six-stage video authoring workflow, with the Seed Panel supporting intent negotiation, a shared AI Agent enabling cross-stage collaboration, and two knowledge bases supporting differentiated intent expansion and faithful translation.}
\label{fig:1}
\Description{System architecture diagram of the proposed video authoring system. Three layers are connected by upward arrows. The top layer shows a six-stage AI Video Authoring Workflow from Input Creative Idea to Editing and Integration, with a labeled workflow bar beneath the stage boxes. The middle layer is the Seed Panel, with a UI mockup showing editable seeds, a preview image, and an AI assistant chat, plus three labeled components: Seed Space, AI Agent Suggestions, and Result Preview; a shared AI Agent bar below connects upward to the Seed Panel and to Editing and Integration. The bottom layer, Knowledge Augmentation, contains the Multi-style Script Library (15 story development structures and 15 narrative styles) and the Cinematic Language Knowledge Base covering focal length and depth of field, camera movement, camera angle, shot size, color, and lighting.}
\end{figure}

Based on these findings, we designed and implemented VideoMaker (Figure~\ref{fig:1}). The system operationalizes creative intent as an evolving shared state across a six-stage video authoring workflow: Input Creative Idea \ensuremath{\rightarrow} Co-create the Script \ensuremath{\rightarrow} Co-create the Storyboard Script \ensuremath{\rightarrow} Define Visual Style and Characters \ensuremath{\rightarrow} Generate Images and Videos \ensuremath{\rightarrow} Editing and Integration. A unified AI Agent operates across the workflow, carrying creative intent across stages and supporting its progressive transformation into concrete representations. During early-stage ideation, the Agent draws on the Multi-style Script Library, which provides alternative Story Development Structures and Narrative Styles to support differentiated expansion of the initial intent. As creation progresses, the Cinematic Language Knowledge Base helps the Agent translate creators’ natural and abstract expressions into structured narratives and executable audiovisual descriptions. At key stages—including script creation, individual storyboard-shot editing, style and character definition, and image/video generation—the Seed Panel externalizes creator-confirmed intent as editable seeds while leaving unspecified space open to AI recombination, supplementation, and proactive suggestions. Creators can continuously confirm, add, remove, and revise these seeds throughout the workflow. This enables what should remain stable and what remains open to AI initiative to be explicitly negotiated as creative intent evolves across stages.\par

We further evaluated VideoMaker through a within-subjects study with 30 participants. Using a mixed-methods evaluation combining think-aloud protocols, questionnaires, and semi-structured interviews, we assessed its support for the three design goals. The results showed that VideoMaker demonstrated significant advantages in the differentiated expansion of initial creative intent and its narrative translation across stages, several dimensions of creative support, and willingness to use, while providing stronger support for preserving creator-confirmed intent. However, negotiated boundaries between creator-confirmed intent and AI initiative remained difficult to sustain across creative stages, suggesting that these boundaries need to be adaptively adjusted and renegotiated according to stage-specific needs.\par

This paper makes the following contributions:\par

\begin{itemize}
\item A conceptualization of creative intent as an evolving shared state. We identify intent expansion, cross-stage translation, and intent negotiation as three interrelated challenges in human–AI video co-creation.

\item A seed-based mechanism for cross-stage intent negotiation. The Seed Panel externalizes creator-confirmed intent as editable, persistent seeds while leaving unspecified space open to AI initiative.

\item Empirical evidence from VideoMaker. A within-subjects study with 30 participants demonstrates benefits for design goals, while revealing limitations in maintaining negotiated intent boundaries across stages.

\end{itemize}

\section{Related Work}
\label{sec:2}

\subsection{Needs and Characteristics of AI-assisted Video Creation}
\label{sec:2.1}

Advances in generative AI have expanded its role in video from media synthesis to content understanding, creative planning, production, editing, and distribution, increasingly spanning the full creation process \cite{zhou2024surveyvideo}. In filmmaking, AI has been incorporated into shot organization, media processing, and post-production \cite{wang2024filmmaking}. In short-form video creation, it supports tasks from topic ideation and scriptwriting to media generation, editing, and publishing \cite{kim2024unlocking,anderson2025making,xia2022synthesis}. Domain-specific systems further integrate professional knowledge into music video, educational video, and news video, lowering barriers to specialized creation \cite{lee2025mvprompt,baek2026pedacogen,wang2024reelframer,xia2022synthesis}. Accordingly, creator–AI interaction is shifting from isolated tool use toward continuous collaboration throughout the creative process.\par

AI-assisted video creation is often conflated with video generation, video editing, and video production. Video generation concerns the technical capability to synthesize and transform audiovisual content \cite{jiang2025vace,ma2025videodiffusion}. Video editing involves operations on existing media, including retrieval, selection, trimming, sequencing, and assembly \cite{tilekbay2024expressedit,wang2024lave,wang2024podreels,wang2026rewriting}. Video production refers to the broader execution pipeline spanning planning, asset acquisition, post-production, and delivery \cite{rabowsky2025video,zhang2025filmcreation,lee2025shortform}. These functions constitute parts of AI-assisted video creation, but do not center on creative intent \cite{kadenhe2025codesign}. AI-assisted video creation, as examined in this work, organizes these distributed activities and outputs around creative goals and narrative structures, while continuously adapting to intermediate results. Its key concern is whether creative intent can continue to shape the evolving work while itself being progressively specified and revised \cite{zhang2025agencyreview,kadenhe2025codesign,rabowsky2025video}.\par

AI-assisted video creation typically involves four recurring processes through which creative intent becomes progressively specified. Intent formation and expansion. Initial ideas are developed into alternative themes, scenes, and narrative relations, with AI-generated possibilities selected and refined into clearer directions \cite{lee2025mvprompt,yeh2026vidmento,li2026creatorcentric}. From ideation to scripts and storyboards. Intent is translated across representations such as natural language, hierarchical documents, story structures, and storyboards into scripts, scenes, shots, and temporal sequences \cite{antony2024id8,liu2026textnative}. Previsualization and generation control. Making spatial layouts, character performance, camera relationships, and audiovisual rhythm inspectable before they are materialized as images, speech, music, and video, enabling repeated comparison and adjustment \cite{hu2026previzwhiz,wang2026script2screen,li2026sketchdynamics}. Iterative refinement. Allowing creators to revisit earlier stages when generated results reveal narrative gaps, stylistic drift, or shot-level problems, using revision, comparison, and local regeneration to refine the work \cite{yeh2026vidmento,li2026creatorcentric,lee2025shortform}. Creators frequently move back and forth among these stages, making AI-assisted video creation a multi-stage, multi-representational, nonlinear, and iterative activity.\par

AI-assisted video creation examined in this work is therefore a highly demanding activity \cite{anderson2025making,lee2025shortform}. Creators must coordinate multiple stages and tools while integrating creative, narrative, and audiovisual expertise \cite{rabowsky2025video,zhang2025agencyreview}. Meanwhile, creative intent continuously develops and carries across these stages and representations, giving rise to substantial challenges.\par

\subsection{Challenges in AI-assisted Video Creation}
\label{sec:2.2}

Research on AI-assisted video creation has grown rapidly, but existing work has largely focused on model capability, generation quality, and tool efficiency \cite{ma2025videodiffusion,ma2025controllable,jiang2025vace}. Systematic discussion centered on creative intent remains limited \cite{zhang2025filmcreation,kim2024unlocking}. The following summarizes the main challenges across stages of the creative process.\par

During intent formation and expansion, creators explore possible directions. Generative AI can rapidly provide references through text and image generation \cite{hoggenmueller2023creative}, yet its outputs are often shaped by learned data patterns, which may lead to design fixation, reduced diversity, and lower originality \cite{wadinambiarachchi2024effects,doshi2024generative,kim2024unlocking,khan2025beyond}. Some systems support exploration through multiple alternatives and comparative selection \cite{huh2025videodiff,liu2026textnative,antony2024id8,yeh2026vidmento}, but these approaches mainly expand choice among generated outcomes rather than the novelty of the content itself. Thus, a key challenge at this stage is how AI can expand creators’ creative space without reinforcing homogenized patterns.\par

During the transition from ideation to scripts and storyboards, creators organize intent into narrative structures. Natural-language descriptions often underspecify audiovisual, spatial, and temporal details, making them difficult to translate directly into executable plans \cite{shin2024lostmelody,liu2026textnative,li2026creatorcentric,wang2026script2screen,hu2026previzwhiz}. Prior work introduces intermediate representations to decompose abstract ideas into more explicit elements \cite{antony2024id8,lee2025mvprompt,wang2024reelframer,baek2026pedacogen}; sketches and other multimodal forms help creators express complex intent \cite{wen2025seewhat,hu2026previzwhiz,li2026sketchdynamics,tilekbay2024expressedit,wang2026script2screen,liu2026textnative}. However, these approaches primarily address how information is structured or expressed. As intent is translated across multiple representations, each step may reinterpret it, creating intent drift between the original idea and the resulting video \cite{shin2024lostmelody,liu2026textnative,li2026creatorcentric,wang2026script2screen,hu2026previzwhiz}.\par

During previsualization and generation control, creators refine their ideas in response to generated results. Video creation requires temporal continuity, spatial consistency, and narrative coherence \cite{ma2025videodiffusion}, while text alone is often insufficient for expressing complex requirements \cite{ma2025controllable}. Existing work improves controllability through conditional generation, multimodal inputs, and interactive editing \cite{jiang2025vace,ma2025videodiffusion,ma2025controllable}; some systems further support local editing, multiple alternatives, and comparison \cite{wang2026rewriting,huh2025videodiff}. Yet more control does not necessarily produce better collaboration: excessive specification can reduce AI to an execution tool, whereas excessive AI autonomy can lead the work away from creative intent. The challenge is therefore not simply to increase control, but to make explicit which parts of creative intent should be confirmed and preserved by the creator and which parts remain open to AI initiative.\par

During iterative refinement, creators repeatedly return to earlier stages. Existing systems support iteration through synchronized editing across representations, versioning, feedback analysis, and reversible operations \cite{antony2024id8,yeh2026vidmento,liu2026textnative,li2026creatorcentric,wang2026script2screen}, while AI can also provide analysis and refinement suggestions \cite{wang2024podreels,wang2024lave,wang2024reelframer,yeh2026vidmento,baek2026pedacogen,savytski2026bridging}. However, most approaches optimize individual stages rather than addressing cross-stage continuity throughout a long creative process \cite{antony2024id8,li2026creatorcentric,yeh2026vidmento}.\par

Overall, the central challenges of AI-assisted video creation can be understood through three dimensions: creative intent must be expanded without being displaced by common generative patterns; translated across stages and representations while maintaining continuity; and negotiated across stages with respect to which aspects should remain creator-confirmed and which should remain open to AI initiative. Therefore, rather than treating creative intent as a fixed input to be statically preserved, we view it as an evolving state that is continuously expanded, translated, and negotiated throughout human–AI co-creation.\par

\subsection{Co-creation Mechanisms in AI-assisted Video Creation}
\label{sec:2.3}

In response to the challenges discussed above, we review prior work from a human–AI co-creation perspective and summarize how existing systems support the evolution of creative intent.\par

\begin{itemize}
\item Multi-Alternative Exploration and Comparison

\end{itemize}

Prior work provides selectable creative resources by proactively generating multiple alternatives and supporting their alignment, comparison, management, and regeneration. Examples include aligning video alternatives around shared anchors and highlighting their differences \cite{huh2025videodiff}, offering multiple variants with flexible human takeover in structured workflows \cite{liu2026textnative,antony2024id8}, and combining context-aware expansion, variant comparison, and local regeneration \cite{yeh2026vidmento}. These mechanisms support divergent exploration, though homogenization of AI-generated content remains a concern \cite{wadinambiarachchi2024effects}.\par

\begin{itemize}
\item Structured Creative Scaffolds and Intermediate Representations

\end{itemize}

Another line of work decomposes vague creative intent into actionable structured elements or intermediate representations, including storylines, scene-by-scene scripts, storyboards, scenes, and timelines \cite{antony2024id8}; visual themes, scene composition, and visual grammar \cite{lee2025mvprompt}; narrative framings and foundational elements \cite{wang2024reelframer}; script blueprints and workflows \cite{baek2026pedacogen}; and compositional substrates spanning spatial, temporal, narrative, and congruent aspects \cite{cao2025compositional}. Such representations reduce the gap between abstract ideas and generated content, allowing creative intent to be progressively specified and inspected.\par

\begin{itemize}
\item Multimodal and Novel Interaction

\end{itemize}

Moving beyond text-only conversational interaction, prior work supports creators’ expression through multimodal inputs and alternative interaction techniques. Examples include real-time VR experiences \cite{wen2025seewhat}, rough 3D scenes \cite{hu2026previzwhiz}, free-form sketches \cite{li2026sketchdynamics}, combined natural-language and sketch input \cite{tilekbay2024expressedit}, and interactive 3D audiovisual previsualization \cite{wang2026script2screen} for expressing complex information such as motion and spatial composition. Other systems convert video into editable text \cite{wang2026rewriting}, augment text editors \cite{lin2026visuallyrics}, or map plain-text documents directly onto video structures \cite{liu2026textnative}, enabling video authoring through new reading- and writing-based interactions. Together, these approaches lower the barrier to expressing complex creative intent and enable more explicit articulation of creators’ ideas.\par

\begin{itemize}
\item Suggestive Assistance and Creator-Led Decision Making

\end{itemize}

Another line of work positions AI as an assistant for content analysis, clip recommendation, review, and feedback, with outputs presented as suggestions that creators can accept, edit, or reject. PodReels and LAVE support footage selection through AI-based content analysis and storyboard planning \cite{wang2024podreels,wang2024lave}. ReelFramer, Vidmento, and PedaCo-Gen adopt suggestive workflows for script generation, narrative expansion, and educational video design, respectively \cite{wang2024reelframer,yeh2026vidmento,baek2026pedacogen}. CHIEF further introduces an iterative feedback loop through persona-conditioned agents \cite{savytski2026bridging}. These systems generally preserve creators’ editing rights and decision-making authority. However, they typically maintain a relatively fixed relationship between human control and AI assistance, with limited consideration of which parts of creative intent should remain creator-confirmed and which remain open to AI initiative~\cite{gao2026hchac}.\par

Viewed across the input–action–output–feedback stages of human–AI co-creation \cite{zhang2025agencyreview}, existing tools have explored structured and multimodal input, multimodal and role-based interaction, multi-alternative comparison of outputs, and iterative feedback loops. However, as summarized in Table~\ref{tab:1}, most approaches address problems at individual stages, with limited consideration of expanding, translating, and negotiating creative intent across stages. Therefore, our work explores a co-creation mechanism that treats creative intent as an evolving shared state, enabling creators and AI to jointly develop and adjust this state throughout the creative process.\par

\begin{table}[!htbp]
\centering
\caption{Comparison of representative AI-assisted video co-creation tools across four core dimensions. \ensuremath{\mdlgblkcircle} = Full support, \ensuremath{\circlelefthalfblack} = Partial support, \ensuremath{\mdlgwhtcircle} = Not supported.}
\label{tab:1}
\Description{Comparison table of ten representative AI-assisted video co-creation tools across four dimensions. Legend: filled circle equals full support, half-filled circle equals partial support, empty circle equals not supported. Columns are Expansion of creative intent, Cross-representational translation, Negotiation between creator and AI, and Cross-stage continuity. ID.8: full, full, partial, not supported. VideOrigami: partial, full, partial, full. CHIEF: full, partial, full, full. Vidmento: full, partial, partial, not supported. Doki: full, partial, partial, partial. PedaCo-Gen: partial, full, full, not supported. ReelFramer: partial, full, partial, not supported. VideoDiff: full, partial, partial, partial. ExpressEdit: full, full, partial, partial. LAVE: full, partial, partial, partial. No tool shows full support on all four dimensions; VideOrigami and CHIEF offer the strongest cross-stage continuity.}
\begingroup
\footnotesize
\setlength{\tabcolsep}{3pt}
\renewcommand{\arraystretch}{1.08}
\begin{tabularx}{\linewidth}{@{}>{\raggedright\arraybackslash\hsize=1.25\hsize}X>{\centering\arraybackslash\hsize=0.9375\hsize}X>{\centering\arraybackslash\hsize=0.9375\hsize}X>{\centering\arraybackslash\hsize=0.9375\hsize}X>{\centering\arraybackslash\hsize=0.9375\hsize}X@{}}
\toprule
Tool & Expansion of creative intent & Cross-representational translation of creative intent & Negotiation between creator and AI & Cross-stage continuity \\
\midrule
ID.8~\cite{antony2024id8} & \ensuremath{\mdlgblkcircle} & \ensuremath{\mdlgblkcircle} & \ensuremath{\circlelefthalfblack} & \ensuremath{\mdlgwhtcircle} \\
VideOrigami~\cite{cao2025compositional} & \ensuremath{\circlelefthalfblack} & \ensuremath{\mdlgblkcircle} & \ensuremath{\circlelefthalfblack} & \ensuremath{\mdlgblkcircle} \\
CHIEF~\cite{savytski2026bridging} & \ensuremath{\mdlgblkcircle} & \ensuremath{\circlelefthalfblack} & \ensuremath{\mdlgblkcircle} & \ensuremath{\mdlgblkcircle} \\
Vidmento~\cite{yeh2026vidmento} & \ensuremath{\mdlgblkcircle} & \ensuremath{\circlelefthalfblack} & \ensuremath{\circlelefthalfblack} & \ensuremath{\mdlgwhtcircle} \\
Doki~\cite{liu2026textnative} & \ensuremath{\mdlgblkcircle} & \ensuremath{\circlelefthalfblack} & \ensuremath{\circlelefthalfblack} & \ensuremath{\circlelefthalfblack} \\
PedaCo-Gen~\cite{baek2026pedacogen} & \ensuremath{\circlelefthalfblack} & \ensuremath{\mdlgblkcircle} & \ensuremath{\mdlgblkcircle} & \ensuremath{\mdlgwhtcircle} \\
ReelFramer~\cite{wang2024reelframer} & \ensuremath{\circlelefthalfblack} & \ensuremath{\mdlgblkcircle} & \ensuremath{\circlelefthalfblack} & \ensuremath{\mdlgwhtcircle} \\
VideoDiff~\cite{huh2025videodiff} & \ensuremath{\mdlgblkcircle} & \ensuremath{\circlelefthalfblack} & \ensuremath{\circlelefthalfblack} & \ensuremath{\circlelefthalfblack} \\
ExpressEdit~\cite{tilekbay2024expressedit} & \ensuremath{\mdlgblkcircle} & \ensuremath{\mdlgblkcircle} & \ensuremath{\circlelefthalfblack} & \ensuremath{\circlelefthalfblack} \\
LAVE~\cite{wang2024lave} & \ensuremath{\mdlgblkcircle} & \ensuremath{\circlelefthalfblack} & \ensuremath{\circlelefthalfblack} & \ensuremath{\circlelefthalfblack} \\
\bottomrule
\end{tabularx}
\endgroup
\end{table}

\section{Formative Study}
\label{sec:3}

We first conducted semi-structured interviews \cite{adams2008questionnaires,kjeldskov2014semistructured,kim2024unlocking} to understand creators’ actual workflows, challenges, and collaboration needs in AI-assisted video creation, with a particular focus on how creative intent evolves throughout a multi-stage creative process with sustained AI involvement. To capture different levels of experience, we recruited both novice and experienced AI creators. We then conducted thematic analysis \cite{braun2006thematic,byrne2022worked,braun2024thematic,kim2024unlocking} to identify key themes and patterns and derive the system design goals. We subsequently invited three experts to participate in a co-design process, translating these goals into concrete design mechanisms.\par

\subsection{Semi-Structured Interviews}
\label{sec:3.1}

We used purposive sampling to recruit participants with varying levels of creative experience. The inclusion criteria were: (1) experience with video creation; (2) prior use of at least one AI image or video generation tool; and (3) willingness to describe their workflows in detail. We recruited 12 participants, including six novice creators (fewer than 10 completed works and less than one year of creation experience) and 6 experienced creators (at least 10 completed works or at least one year of creation experience), labeled C01–C12. Participants were recruited through creative communities, social media, and peer referrals. Their demographics are summarized in Table~\ref{tab:2}. The sample size was consistent with qualitative research practice, and thematic saturation was reached.\par

\begin{table}[!htbp]
\centering
\caption{Demographics and AI usage of participants in the formative study.}
\label{tab:2}
\Description{Demographics and AI usage for 12 formative-study participants (C01-C12). Columns: ID, Gender, Age, Creation Type, Group, AI Use Frequency. C01: Female, 46, Short-form videos, Novice, Sometimes. C02: Male, 18-25, Short-form videos, Novice, Sometimes. C03: Male, 18-25, Short-form videos, Novice, Sometimes. C04: Male, 18-25, Short-form videos, Novice, Sometimes. C05: Female, 37, IP-based animated videos, Novice, Sometimes. C06: Male, 18-25, Commercial videos, Experienced, Frequently. C07: Female, 18-25, AI-generated short-form videos, Experienced, Frequently. C08: Female, 18-25, AI-generated short-form videos, Experienced, Frequently. C09: Female, 18-25, AI-generated short-form videos, Novice, Rarely. C10: Male, 18-25, Videos, Experienced, Frequently. C11: Male, 18-25, Short-form videos, Experienced, Frequently. C12: Female, 18-25, AI-generated short-form videos, Experienced, Frequently. Six novices and six experienced creators; AI use ranges from rarely to frequently.}
\begingroup
\footnotesize
\setlength{\tabcolsep}{3pt}
\renewcommand{\arraystretch}{1.08}
\begin{tabularx}{\linewidth}{@{}>{\raggedright\arraybackslash\hsize=0.6192\hsize}X>{\raggedright\arraybackslash\hsize=0.8161\hsize}X>{\raggedright\arraybackslash\hsize=0.7584\hsize}X>{\raggedright\arraybackslash\hsize=1.6352\hsize}X>{\raggedright\arraybackslash\hsize=1.0391\hsize}X>{\raggedright\arraybackslash\hsize=1.1320\hsize}X@{}}
\toprule
ID & Gender & Age & Creation Type & Group & AI Use Frequency \\
\midrule
C01 & Female & 46 & Short-form videos & Novice & Sometimes \\
C02 & Male & 18–25 & Short-form videos & Novice & Sometimes \\
C03 & Male & 18–25 & Short-form videos & Novice & Sometimes \\
C04 & Male & 18–25 & Short-form videos & Novice & Sometimes \\
C05 & Female & 37 & IP-based animated videos & Novice & Sometimes \\
C06 & Male & 18–25 & Commercial videos & Experienced & Frequently \\
C07 & Female & 18–25 & AI-generated short-form videos & Experienced & Frequently \\
C08 & Female & 18–25 & AI-generated short-form videos & Experienced & Frequently \\
C09 & Female & 18–25 & AI-generated short-form videos & Novice & Rarely \\
C10 & Male & 18–25 & Videos & Experienced & Frequently \\
C11 & Male & 18–25 & Short-form videos & Experienced & Frequently \\
C12 & Female & 18–25 & AI-generated short-form videos & Experienced & Frequently \\
\bottomrule
\end{tabularx}
\endgroup
\end{table}

IRB approval was obtained, and all participants provided informed consent before the interviews. Interviews were conducted online or in person (40–60 minutes). With consent, all interviews were recorded, professionally transcribed, double-checked by the first two authors, and anonymized.\par

The interviews began with participants’ actual experiences of AI-assisted video creation and covered: (1) background and experience; (2) existing AI-assisted video creation practices and workflows; (3) needs and challenges throughout the creative process; and (4) expectations for human–AI collaboration; see Appendix~\ref{sec:A}. To account for differences in technical experience, we recorded participants’ prior use of AI tools and briefly introduced the basic capabilities of current AI video creation tools.\par

We conducted an inductive thematic analysis \cite{braun2006thematic,byrne2022worked,braun2024thematic,kim2024unlocking}. Two researchers independently coded approximately 20\% of the data to identify behaviors, challenges, and needs relevant to the research questions. They then discussed code names, definitions, and disagreements, and iteratively refined the codebook. The remaining data were coded, after which related codes were grouped into broader categories and synthesized into higher-level themes.\par

\subsection{Interview Findings}
\label{sec:3.2}

Our formative study showed that creators expect AI to provide sustained support throughout video creation, from ideation to iterative refinement. The core tension is how creative intent can be constructed as an evolving shared state. This question manifests in three dimensions—intent expansion, translation, and negotiation—which informed the three design goals shown in Table~\ref{tab:3}.\par

\begin{table}[!htbp]
\centering
\caption{Mapping creators' needs to design goals based on the semi-structured interviews.}
\label{tab:3}
\Description{Mapping table from formative interview needs to three design goals. Four columns: Need, Tension, Design Question, Design Goal. Row 1: Need, intent should remain differentiated. Tension, AI alternatives cluster around common narrative patterns. Design question, can AI expand creative intent rather than replace it with common patterns. Design goal DG1, support differentiated expansion around creative intent. Row 2: Need, intent should carry through the process. Tension, creators struggle to translate intent and AI may cause intent drift across representations. Design question, can AI support stable translation rather than repeated reinterpretation. Design goal DG2, support progressive and faithful translation across stages and representations. Row 3: Need, key creative decisions should remain with the creator while collaborating with AI. Tension, excessive specification or AI autonomy can weaken collaboration. Design question, can creators and AI negotiate what to preserve versus what remains open. Design goal DG3, support continuously negotiable boundaries between creator-confirmed intent and AI initiative.}
\begingroup
\footnotesize
\setlength{\tabcolsep}{3pt}
\renewcommand{\arraystretch}{1.08}
\begin{tabularx}{\linewidth}{@{}>{\raggedright\arraybackslash\hsize=0.6237\hsize}X>{\raggedright\arraybackslash\hsize=1.4846\hsize}X>{\raggedright\arraybackslash\hsize=1.0592\hsize}X>{\raggedright\arraybackslash\hsize=0.8325\hsize}X@{}}
\toprule
Need & Tension & Design Question & Design Goal \\
\midrule
My intent should remain differentiated. & As discussed in Sections~\ref{sec:2.2} and 3.2.1, AI can rapidly generate many ideas, but these alternatives often cluster around common narrative patterns. Creators may therefore end up choosing within a limited space predefined by AI. & Can AI help creators expand creative intent, rather than replacing the current direction with common AI-generated patterns? & DG1. Support differentiated expansion around creative intent. \\
\addlinespace[2.5pt]
My intent should carry through the process. & As discussed in Sections~\ref{sec:2.2}, 3.2.2, 3.2.3 and 3.2.4, creators often “know what they want” but struggle to translate it into professional expression. Their intent must also undergo repeated translation across representations and stages, during which AI may reinterpret or supplement it, causing intent drift. & Can AI support the stable translation of creative intent, rather than repeatedly reinterpreting it during successive translation and generation? & DG2. Support the progressive and faithful translation of creative intent across stages and representations. \\
\addlinespace[2.5pt]
Key creative decisions should remain mine while I collaborate with AI. & As discussed in Sections~\ref{sec:2.2}, 3.2.3, 3.2.4, and 3.2.5, creators want both local control and sustained collaboration with AI. Excessive human specification restricts AI’s generative space, whereas excessive AI autonomy weakens creator-led decision making. Their preferred degree of AI initiative also varies across creators and creative states. & Can creators and AI explicitly and continuously negotiate which parts of the current creative intent should be preserved, and which parts remain open for AI to explore? & DG3. Support continuously negotiable boundaries between creator-confirmed intent and AI initiative. \\
\bottomrule
\end{tabularx}
\endgroup
\end{table}

\subsubsection{Difficulty in Differentiated Expansion of Creative and Narrative Directions (11/12)}
\label{sec:3.2.1}

During topic selection and ideation, the main difficulty was developing creative directions that avoid conventional patterns. Many creators expressed fatigue with formulaic promotional videos and sought entry points distinct from common narrative conventions (C02, C06, C07). For example, C06 wanted to avoid “formulaic PPT/MV-style promotional videos.” When working with broad or abstract themes, participants often “knew what to make but not how to make it novel,” and relied on external references or AI for inspiration (C03, C07). More AI-generated ideas did not necessarily expand creators’ own creative space; some participants worried that common AI-generated patterns could “pull the original direction off course” (C04, C05).\par

This difficulty extended to narrative organization. When translating abstract ideas into stories, participants struggled with sequencing complex backgrounds, establishing causal relations between events, and maintaining a clear narrative thread, while AI often suggested “formulaic” narrative directions (C01, C03, C09, C11, C12). These findings indicate a need for differentiated directions that expand from creators' current creative intent and remain open to further development.\par

\subsubsection{Barriers from Creative Intent to Professional Audiovisual Expression (10/12)}
\label{sec:3.2.2}

Creators commonly faced the problem of “knowing what they want, but struggling to express it precisely.” Participants often lacked the expertise to translate visual intentions into professional terms such as shot size and composition (C03, C05, C12). C11, for example, found it difficult to “translate an artistic conception into visual descriptions” because “visual language and textual language are different.” Unfamiliarity with professional terminology often led to missing prompt information and unexpected outputs (C04, C07). Shot specifications, environmental details, and background information were also easily omitted, while the lack of visual examples made their effects harder to understand (C09). This problem extended beyond individual shots. Participants also lacked systematic guidance on shot transitions, camera movements, and continuity across shot groups (C10, C11). Many therefore expected lower barriers to professional terminology and more visual support for understanding cinematic concepts (C11, C12).\par

\subsubsection{Intent Drift in Generation and the Need for Local Refinement (11/12)}
\label{sec:3.2.3}

The randomness of AI generation imposed high trial-and-error costs, with abrupt transitions, stylistic discontinuities, and character clipping recurring across outputs (C06, C08). Creators strongly resisted regenerating entire results. Instead, they wanted videos decomposed into independent shots so that local details could be refined while satisfactory segments were preserved (C04, C07, C08, C09). C07, for example, expected the Agent to detect and correct problems such as “shot clipping” and “logical inconsistencies.” Participants also wanted to adjust content during generation, exert finer control over camera movement and image-to-video generation (C10), and verify results through previews (C09, C11). When scripts lacked sufficient detail, they expected more specific descriptions, comparisons across multiple versions, and confirmation of relationships between shots (C11, C12).\par

\subsubsection{Need for Cross-Stage Intent Continuity (9/12)}
\label{sec:3.2.4}

Participants generally expected AI to act as a persistent collaborator that understands and carries forward established creative intent and content across stages. C08 wanted an Agent that connected the entire creative process so that characters, visual styles, and narrative information established earlier could naturally carry into subsequent generation. C10 suggested first organizing shots into groups and then having AI maintain continuity within each group. C09 proposed using a unified style reference so that later generations could follow an already confirmed visual direction. This need also appeared in participants’ desire to see how AI interprets their current intent earlier and revise it before moving to the next stage. For example, they wanted previews during scriptwriting so they could evaluate how the textual narrative would translate visually before proceeding further (C02, C07, C11). These findings suggest that creative intent must not only be translated within individual stages, but also carried forward as a persistent state across subsequent stages.\par

\subsubsection{Diverse Needs for Collaboration Modes and Decision Boundaries (12/12)}
\label{sec:3.2.5}

Participants generally wanted AI to participate more deeply in video creation, but differed substantially in how proactive AI should be. When they lacked a clear direction, some wanted the Agent to act as a guide, asking questions to fill in missing creative details (C01, C03). More experienced creators preferred a “senior assistant director” that could proactively identify narrative or shot-level problems and offer professional suggestions (C06, C08). Rather than a fixed division of labor, participants therefore expected AI involvement to adapt to their experience and current creative state.\par

Preferences for interaction also varied. Participants suggested voice interaction (C01), visual block-based manipulation (C05), storyboard sketches (C07), and “presets + annotations” (C09). C10 argued that the level of AI assistance should vary with the creator’s familiarity with the topic, while C11 and C12 emphasized item-level annotations, visual sketches, and feedback on motion completeness.\par

These differences reflect a deeper issue of creative decision boundaries. Participants wanted AI to proactively identify problems and provide suggestions beyond their own expertise, while still retaining opportunities to judge and revise important creative directions. The key issue is therefore how creators and AI can continuously negotiate which parts of creative intent should remain creator-confirmed and which remain open to AI initiative.\par

\subsection{Co-Design with Experts}
\label{sec:3.3}

Two of the three experts had extensive frontline experience in video content planning and production; the other focused on the practical application, market dissemination, and commercialization of AI-generated video content. The co-design session centered on interpreting the identified needs, brainstorming solutions, and developing concrete functional mechanisms.\par

\subsubsection{From Homogeneous Alternatives to Differentiated Intent Expansion (DG1)}
\label{sec:3.3.1}

The experts noted that AI-generated creative suggestions in practice are often highly homogeneous, largely recombining existing patterns rather than providing genuinely differentiated inspiration. This observation aligned closely with prior research (Section~\ref{sec:2.2}) and the needs identified in our interviews (Section~\ref{sec:3.2.1}). When discussing where AI should intervene to support ideation, the experts emphasized that although creative stimulation is useful throughout the process, differentiation is often determined earlier, when an initial intent is expanded into distinct story directions. Supporting creators in exploring alternative story development pathways at this stage was therefore considered critical to differentiated expansion. We narrowed the intervention focus of DG1 to the expansion of initial intent.\par

Based on these discussions, we developed a Multi-style Script Library (Section~\ref{sec:4.1}) as a key mechanism for DG1. The library collects script examples with diverse story development structures and narrative styles, providing AI with varied references for expanding the same initial intent.\par

\subsubsection{Progressive and Faithful Translation Inspired by Professional Filmmaking Practices (DG2)}
\label{sec:3.3.2}

The experts argued that professional filmmaking practices and conventions could help reduce reinterpretation across representations and stages. Based on this consensus, we jointly reviewed professional references including \textit{Grammar of the Shot} \cite{thompson2013grammarshot}, \textit{Animation Storyboard Design} \cite{chen2022animation}, \textit{The Visual Story} \cite{block2020visual}, \textit{Directing the Story} \cite{glebas2008directing}, and \textit{Grammar of the Film Language} \cite{arijon1991grammar}. We synthesized these practices into two categories: digital production practices, represented by animation and animated films, and live-action production practices, represented by film and television.\par

Table~\ref{tab:4} summarizes their main differences. In general, digital production offers greater visual freedom but relies heavily on front-loaded planning. Scripts, storyboards, layouts, and shot design progressively specify subsequent production, creating a close correspondence between final assets and earlier plans \cite{chen2022animation,glebas2008directing}. Live-action production is constrained by physical locations and production conditions, but typically retains room for on-set adjustment within an established script and captures additional footage to preserve flexibility in post-production \cite{thompson2013grammarshot}.\par

\begin{table}[!htbp]
\centering
\caption{Comparison between digital and live-action production practices.}
\label{tab:4}
\Description{Comparison table of digital production versus live-action production across five dimensions. Production Medium: digital uses drawing, modeling, and digital techniques; live-action uses camera-based capture. Shot Realization: digital supports difficult viewpoints and spatial effects; live-action is constrained by locations, physics, equipment, and crew. Artistic Characteristics: digital offers high visual freedom and stylized expression; live-action emphasizes realism of people, environments, and events. Expressive Range: digital supports surreal and fantastical digitally constructed scenes; live-action relies on physically available subjects and environments. Creative Workflow: digital uses front-loaded planning from script through shot design with little excess material; live-action follows script-guided on-set adjustment, captures extra footage for post-production, and allows a higher shooting ratio.}
\begingroup
\footnotesize
\setlength{\tabcolsep}{3pt}
\renewcommand{\arraystretch}{1.08}
\begin{tabularx}{\linewidth}{@{}>{\raggedright\arraybackslash\hsize=0.5457\hsize}X>{\raggedright\arraybackslash\hsize=1.2333\hsize}X>{\raggedright\arraybackslash\hsize=1.2211\hsize}X@{}}
\toprule
 & Digital Production & Live-Action Production \\
\midrule
Production Medium & • Drawing and modeling\newline{}• Other digital content-production techniques & • Camera-based image capture\newline{}• Physical recording equipment \\
\addlinespace[2.5pt]
Shot Realization & • Supports conventional cinematography\newline{}• Enables otherwise difficult viewpoints, movements, and spatial effects & • Constrained by locations and physical conditions\newline{}• Limited by equipment and crew coordination \\
\addlinespace[2.5pt]
Artistic Characteristics & • High visual freedom\newline{}• Well suited to stylized and exaggerated expression & • Strong sense of realism\newline{}• Well suited to real people, environments, and events \\
\addlinespace[2.5pt]
Expressive Range & • Digitally constructed environments\newline{}• Supports surreal and fantastical scenes & • Relies on physically available or constructible subjects and environments \\
\addlinespace[2.5pt]
Creative Workflow & • Strongly front-loaded planning\newline{}• Script \ensuremath{\rightarrow} storyboard \ensuremath{\rightarrow} layout \ensuremath{\rightarrow} shot design\newline{}• Final assets closely follow pre-production plans\newline{}• Little or no excess material & • Script- and shooting-plan-guided production\newline{}• Allows on-set adjustment\newline{}• Captures additional footage for post-production flexibility\newline{}• Allows a higher shooting ratio \\
\bottomrule
\end{tabularx}
\endgroup
\end{table}

Because AI video creation involves stochastic generation and frequent consistency challenges \cite{wang2024filmmaking}, we adopted the digital-production logic of front-loaded planning, progressive specification, and stage-wise confirmation as our primary reference, while incorporating selected practices from live-action production. This led to two design mechanisms supporting DG2. First, the AI Video Authoring Workflow (Section~\ref{sec:4.5}) supports the progressive cross-stage translation of creative intent from abstract ideas to final video. Second, the Cinematic Language Knowledge Base (Section~\ref{sec:4.2}) helps translate creators’ vague expressions into professional and executable audiovisual descriptions. Together, they support the progressive and faithful translation of creative intent across multiple stages and representations as it becomes increasingly specified.\par

\subsubsection{Externalizing and Negotiating Creative Intent Through Seed-Based Interaction (DG3)}
\label{sec:3.3.3}

We further examine the interaction modalities between creators and AI, as well as which information should be persistently retained and which should remain open throughout the creative process. Collaborative designers observed that current interactions are predominantly dialogue-based, in which creators’ inputs frequently intermingle core creative intents they wish to maintain continuously with open-ended information intended merely for inspiration. For instance, “the protagonist must re-understand music after losing the sense of hearing” constitutes a critical setting the creator seeks to preserve as an enduring commitment, whereas “one may reference the spatial atmosphere of \textit{Blade Runner 2049}” functions more as an open reference. These two categories of information lack explicit differentiation within extended dialogues; compounded by the frequent occurrence of ambiguous anaphora in natural conversation (e.g., “continue with the previous approach”), this makes it difficult for AI to gauge the current degree of constraint attached to any given piece of information. Over-emphasizing auxiliary information risks constraining AI’s generative space, while neglecting core intent may induce intent drift. Concurrently, a dialogue mode dominated by creator-initiated instructions also limits opportunities for AI to proactively offer inspiration and candidate directions.\par

Collaborative designers contend that the key does not lie in predetermining fixed human–AI divisions of labor, but rather in establishing an adjustable boundary of agency that can be continuously recalibrated during the creative process: creators can explicitly designate which information has become a creative commitment requiring persistent retention, while leaving still-undefined portions open for AI exploration; AI, in turn, can proactively propose novel possibilities within the open space and, upon receiving creator confirmation, incorporate them into the continuously evolving creative intent.\par

To address this issue, we drew on three theoretical perspectives in our co-design:\par

\begin{itemize}
\item Grounding in communication \cite{clark1991grounding} and external cognition \cite{scaife1996external} inspired us to externalize key creative intent from lengthy conversations into a shared representation visible to both creators and AI.

\item Combinational creativity \cite{boden2004creative} and conceptual blending \cite{fauconnier2002way} suggest that new ideas can emerge by selecting, combining, and elaborating elements.

\item Mixed-initiative interfaces \cite{horvitz1999mixedinitiative} inspire us to establish a dynamically adjustable relationship between creator confirmation and AI’s proactive participation.

\end{itemize}

Based on the above discussion, we propose the Seed Panel (see Section~\ref{sec:4.4}), which consists of a Seed Space (composed of multiple Seeds) and an AI Agent. A Seed is an intent that has been explicitly confirmed by the creator and is intended to be persistently retained; it serves as an explicit representation of intent shared between human and AI. Specifically, a Seed constitutes a persistent creative commitment within the current intent state. The Seed Space functions as a discrete creative space that stores these Seeds. Within it, the AI can explore different combinatorial relationships while preserving existing Seeds and proactively propose new candidate Seeds. These candidate Seeds do not automatically become part of the creator’s intent; only after the creator accepts, modifies, or rejects them may they enter the ongoing intent state. The AI Agent is responsible for sustaining dialogue, handling the current intent state across stages, and actively suggesting candidate directions.\par

In this way, creators maintain creator-led decision making by determining “which information must be retained,” while the AI retains its capacity for proactive exploration within spaces that have not been explicitly constrained. As the creative process advances, creators can add, modify, or remove seeds, and the AI can propose new candidate seeds for the open space. Consequently, a continuously negotiable boundary is formed between the intents confirmed by the creator and AI initiative.\par

\section{VideoMaker System}
\label{sec:4}

VideoMaker is an AI-assisted video co-creation system that operationalizes creative intent as an evolving shared state across a six-stage creative workflow. Creators develop a video from an initial idea through six stages: Input Creative Idea, Co-create the Script, Co-create the Storyboard Script, Define Visual Style and Characters, Generate Images and Videos, and Editing and Integration (DG2). At key stages, creators interact with the Seed Panel (DG3) to explicitly confirm, revise, and update the current creative intent. The Seed Panel is integrated with the same AI Agent operating throughout the system, which can access the Multi-style Script Library (DG1) and the Cinematic Language Knowledge Base (DG2) to proactively provide differentiated and domain-informed suggestions. Together, these mechanisms support the expansion, translation, and negotiation of creative intent throughout the workflow.\par

VideoMaker uses FastAPI \cite{ramirez2018fastapi} for the backend and React \cite{react2023documentation}, TypeScript \cite{bierman2014typescript}, and Vite for the frontend, communicating through REST APIs \cite{fielding2000architectural}. Zustand manages cross-page state, while project data is stored in SQLite \cite{gaffney2022sqlite} and creative snapshots are saved as JSON \cite{bray2017json}. The system connects to Qwen3.6-35B-A3B \cite{yang2025qwen3} through an OpenAI-compatible API. Images, videos, and voiceovers are generated using Agnes Image 2.1 Flash, Agnes Video V2.0, and Qwen3-TTS-Flash \cite{hu2026qwen3tts}, respectively.\par

The system includes three key technical modules: (1) Cross-stage Agent context construction, which assembles seeds, creator input, the latest script, storyboards, and other stage-specific information so that the Agent can access and carry forward the current project state across stages; (2) a modular prompt system, with stage-specific prompts covering the workflow from idea decomposition to video editing; and (3) knowledge augmentation, which retrieves relevant content from the Multi-style Script Library and Cinematic Language Knowledge Base according to the current task.\par

The following sections introduce VideoMaker from its underlying knowledge resources to the upper-level workflow. We use the story \textit{Before the Silence} as a running example, beginning with the initial intent: a young composer discovers a box of cassette tapes containing fragmented audio after his mother’s death.\par

\subsection{Multi-style Script Library for Differentiated Expansion of Initial Intent}
\label{sec:4.1}

The Multi-style Script Library extracts two dimensions from representative films and videos: story development structures, which define how plots unfold, and narrative styles, which define how information is organized and presented to the audience. The library contains 15 story development structures, such as the Hero’s Journey, Three-Act Structure, and Road Movie, and 15 narrative styles, including flashback, nonlinear narration, and fragmented narration. Each type is illustrated with 2–5 canonical works, yielding 87 linked cases in total. For example, the Road Movie category includes \textit{Green Book}, \textit{Little Miss Sunshine}, and \textit{Thelma \& Louise}. Each type is further annotated with its definition, emotional tone, pacing, information density, suitable themes, and other retrieval tags.\par

The AI Agent can use these references to produce differentiated adaptations of the same initial intent in \textit{Before the Silence}. For example, using a road-movie structure with flashbacks, the protagonist follows environmental sounds preserved in the recordings to places where his mother once lived. Childhood memories are interwoven throughout the journey, leading him to understand the farewell she never expressed.\par

\subsection{Cinematic Language Knowledge Base for Faithful Translation of Creative Intent}
\label{sec:4.2}

\begin{figure}[!htbp]
\centering
\includegraphics[width=\linewidth]{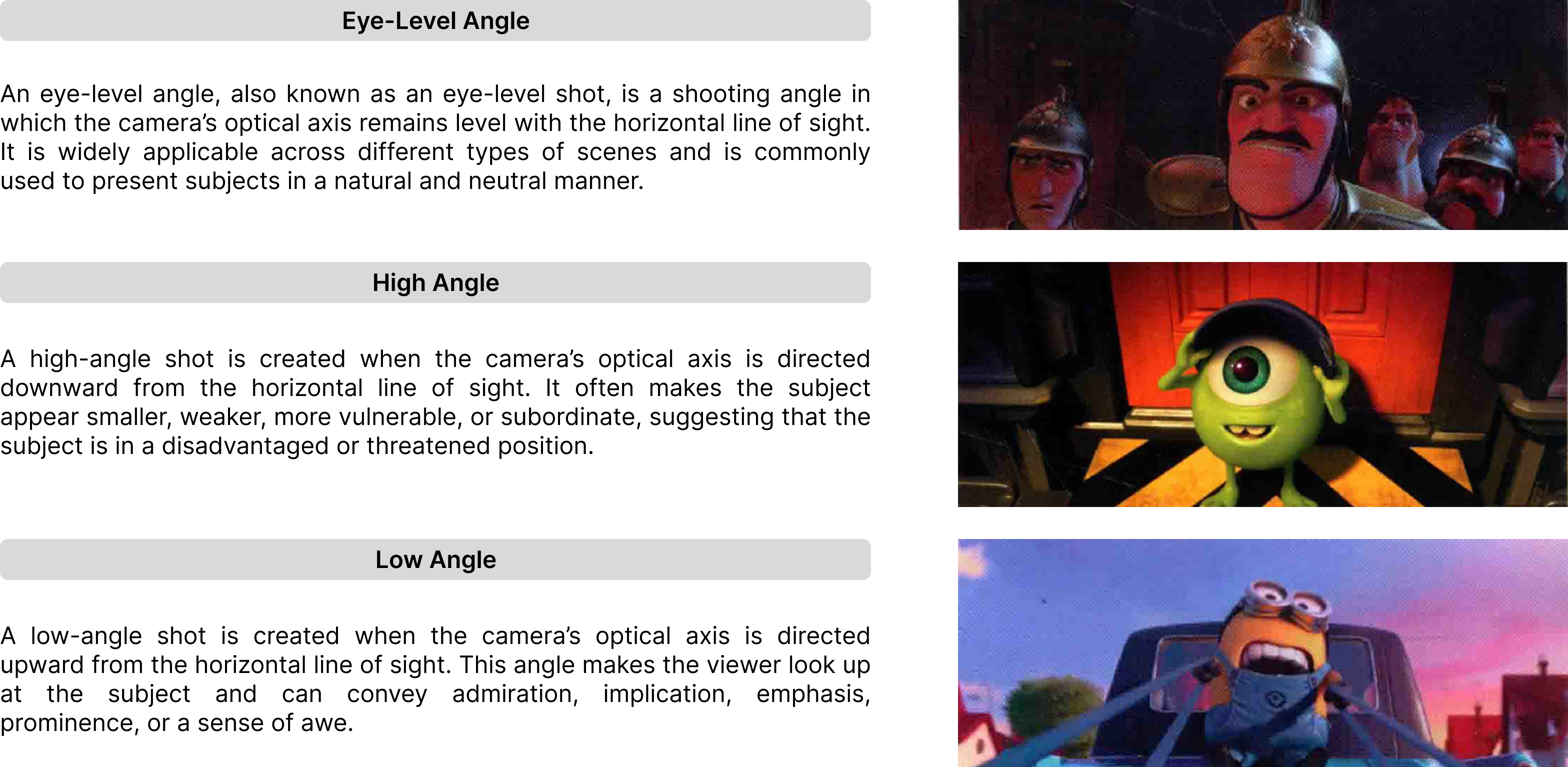}
\caption{Definitions, visual characteristics, effects, and reference images for three camera angles. Reference frames are from \textit{Enchanted} (2007), \textit{Monsters University} (2013), and \textit{Despicable Me 2} (2013), from top to bottom.}
\label{fig:2}
\Description{Three camera-angle entries from the Cinematic Language Knowledge Base. Each row labels one angle and shows a reference still with definition, visual characteristics, and effects. Eye-Level Angle: characters viewed at eye height with a neutral, direct perspective; reference still shows two animated characters facing each other. High Angle: the camera looks down on a small green character in a doorway, making the subject appear smaller or more vulnerable. Low Angle: the camera looks up at yellow minion characters on a vehicle against the sky, making the subjects appear larger and more imposing.}
\end{figure}

The Cinematic Language Knowledge Base covers six categories: focal length and depth of field, shot size, camera angle, camera movement, color, and lighting, comprising 39 reusable cinematic-language entries. For example, the camera-angle category includes eye-level, low-angle, and high-angle shots. Each entry follows a unified structure of definition–visual characteristics and effects–reference image: the definition explains the professional meaning; visual characteristics describe how it appears on screen; effects specify its functions, such as shaping emotion; and reference images help understand the resulting visual effect. These entries serve as intermediate scaffolds between natural-language creative intent and professional audiovisual expression, helping AI supplement the cinematic information required during translation. Figure~\ref{fig:2} shows an example entry for camera angles.\par

\begin{figure}[!htbp]
\centering
\includegraphics[width=\linewidth]{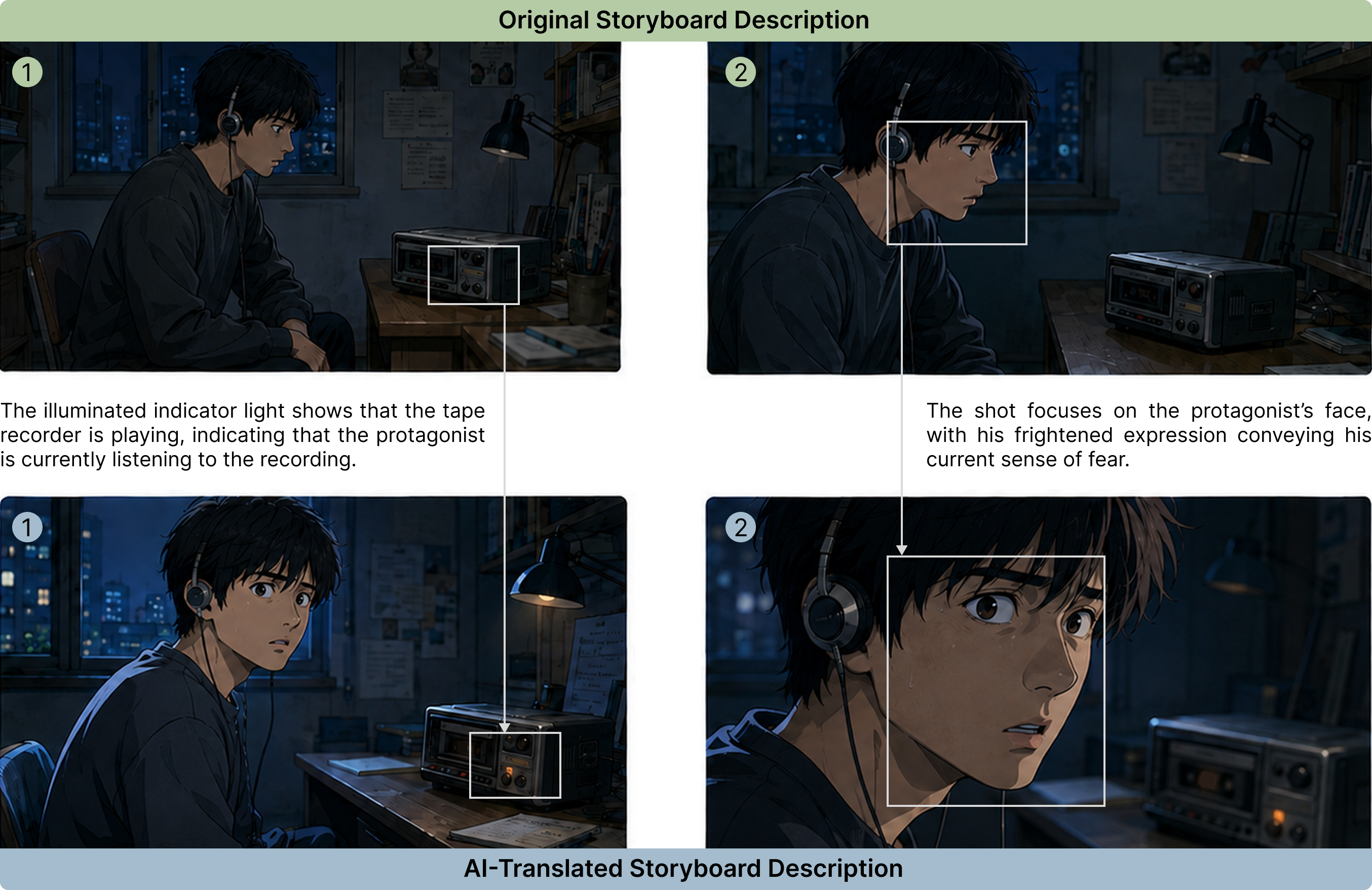}
\caption{Comparison between the creator’s original storyboard description and the description translated by the AI Agent. After AI translation, abstract expressions such as “listens to the tape” and “feels a little scared” are translated into executable audiovisual descriptions, such as the illuminated indicator light on the old tape recorder and a close-up of the protagonist’s frightened and hesitant expression, making the intended visual presentation more explicit.}
\label{fig:3}
\Description{Storyboard comparison with two rows and two columns. The top row, Original Storyboard Description, shows a young man with headphones at a desk at night; panel 1 highlights the cassette player and panel 2 is a neutral close-up of his face. The bottom row, AI-Translated Storyboard Description, uses the same layout with stronger emotion: panel 1 adds a glowing indicator on the cassette player and a startled expression; panel 2 is an extreme close-up with wide eyes and sweat. White boxes and arrows in each row link the cassette-player detail in panel 1 to the facial reaction in panel 2, showing how AI translation shifts focus from object to emotional response.}
\end{figure}

For instance, a creator may describe a storyboard shot as: “The protagonist listens to the tape in his room and feels a little scared. The camera slowly moves closer, and the image should feel colder.” Drawing on the Cinematic Language Knowledge Base, the AI Agent translates these relatively abstract affective, camera, and visual requirements into a more concrete and executable audiovisual description: “The protagonist sits at a desk in his room. The indicator light of an old tape recorder creates a faint warm local light source. The shot begins as an eye-level close shot, with the camera slowly pushing forward into a facial close-up and holding on the young man’s frightened and hesitant expression. The scene uses a low-saturation blue-gray palette.” Figure~\ref{fig:3} shows two sets of storyboard shots generated from these descriptions.\par

\subsection{Cross-stage AI Agent for Maintaining Intent Continuity}
\label{sec:4.3}

The AI Agent serves as the coordination mechanism that maintains creative intent continuity throughout the VideoMaker workflow. The system constructs a continuously updated project context, allowing the Agent to access and carry forward information accumulated throughout the creative process.\par

At each stage, the Agent dynamically integrates creator inputs, confirmed seeds, previous-stage outputs, current scripts and storyboard information, and relevant domain knowledge according to the current task. As creators revise their creative intent, updated information is incorporated into the project context and influences subsequent generation and suggestions.\par

To accommodate different creative stages, the Agent adopts a modular prompt framework (Appendix~\ref{sec:B}) that guides stage-specific tasks. The Agent can also retrieve relevant knowledge from the Multi-style Script Library and Cinematic Language Knowledge Base to provide narrative and cinematic support, enabling creative intent to continuously evolve throughout the workflow.\par

\subsection{Seed Panel: Continuously Negotiating Creator-Confirmed Intent and AI Initiative}
\label{sec:4.4}

\begin{figure}[!htbp]
\centering
\includegraphics[width=\linewidth,height=0.58\textheight,keepaspectratio]{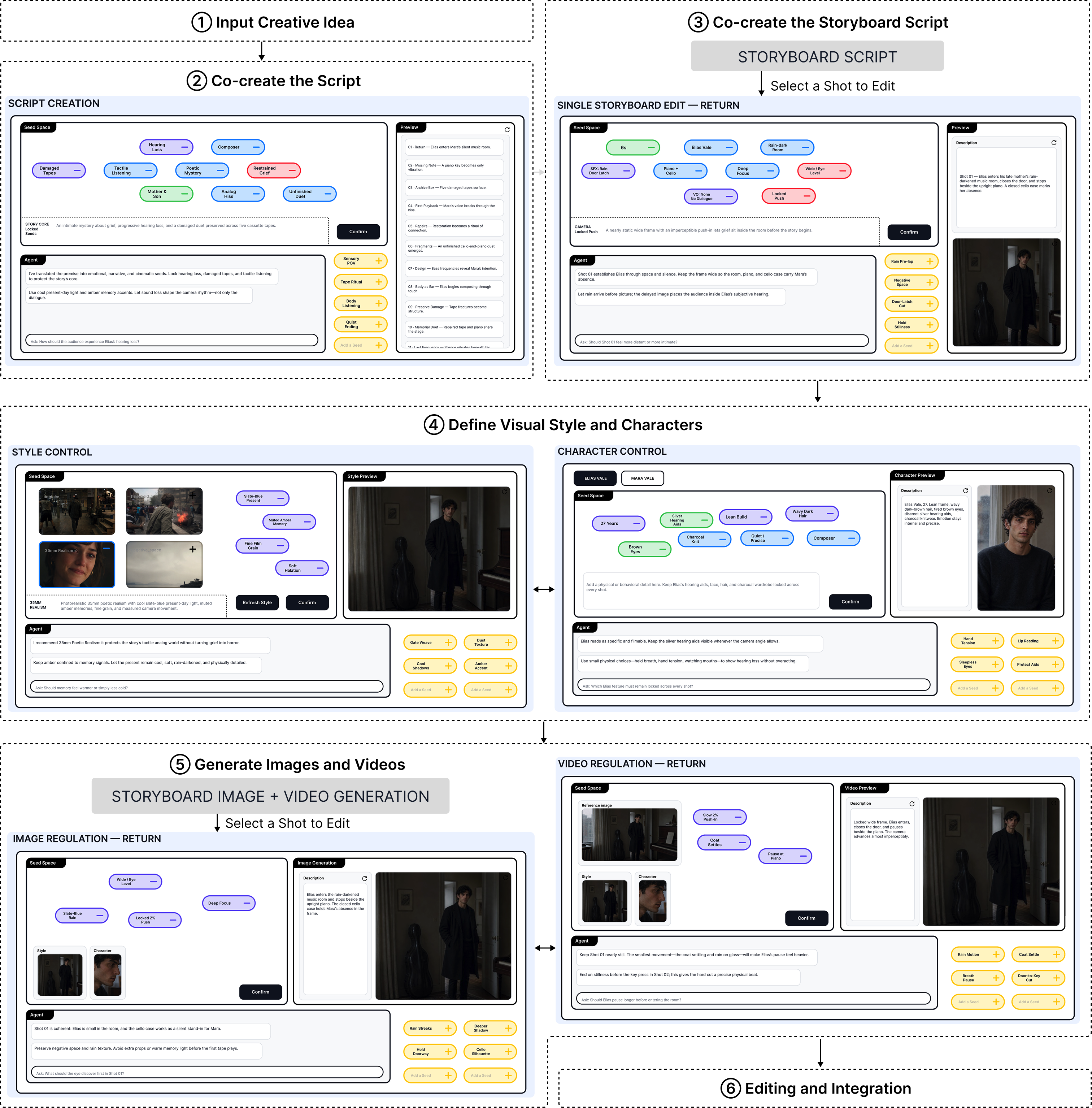}
\caption{Six-stage AI Video Authoring Workflow, in which creative intent is progressively translated across stages and represented as an editable shared state through the Seed Panel in Stages 2–5.\newline{}(1) Input Creative Idea: The creator enters an initial idea. The AI Agent parses the input and recommends story development structures and narrative styles using the Multi-style Script Library. The system then decomposes the confirmed information into initial seeds for script creation.\newline{}(2) Co-create the Script: The creator revises seeds to adjust story development. The AI Agent provides suggestions and generates the script after the creator confirms the seeds.\newline{}(3) Co-create the Storyboard Script: The AI Agent converts the script into a structured storyboard script and supplements each shot with 8 types of information: visual content, voiceover, sound effects, duration, focal length and depth of field, shot size, camera angle, and camera movement. Each shot has its own storyboard seeds, which can be locally edited on the individual-shot editing page.\newline{}(4) Define Visual Style and Characters: The AI Agent identifies key characters from the storyboard, generates character seeds based on their traits, and recommends an overall visual style. Once confirmed, character and style previews serve as consistency constraints for later generations.\newline{}(5) Generate Images and Videos: The AI Agent first identifies the characters involved in each shot, then combines the storyboard script, character seeds, and style seeds to generate image prompts. Creators preview the resulting storyboard images and can continue revising seeds. For video generation, the system extracts motion-related information from the storyboard script to construct prompts and generate video clips.\newline{}(6) Editing and Integration: Based on the storyboard script and generated clips, the AI Agent recommends voiceover, music, and transitions. Creators can revise these elements on the timeline and export the final video editing project file.}
\label{fig:4}
\Description{Six-panel figure of the Seed Panel interface across workflow stages. The figure has two sections. The top section, Storyboard Script, contains four panels. (1) Script Creation: thematic seed tags such as Hearing Loss, Tactile Listening, and Restrained Grief; a Story Core text box; an Agent dialogue area; and a Preview list of numbered story beats including Return and Missing Notes. (2) Single Storyboard Edit—Return: cinematography and sound seeds including 6s duration, Wide/Eye Level, and Piano plus Cello; Agent text for shot 01; Preview with shot description and a thumbnail of a man in a dim room. (3) Style Control: style reference images and tags such as Slate-Blue Present and Fine Film Grain; Agent recommends 35mm Poetic Realism. (4) Character Control: character seeds for Elias Vale including 27 Years, Silver Hearing Aids, and Lean Build; Character Preview shows a portrait of a young man with dark wavy hair. The bottom section, Storyboard Image plus Video Generation, contains two panels. (5) Image Regulation—Return: composition seeds, style and character thumbnails, and an Image Generation preview of a wide shot of a man in a dark room beside a piano and cello case. (6) Video Regulation—Return: motion seeds such as Slow 1\% Push-in and Coat Settles; Agent text on subtle movement; Video Preview shows the rendered cinematic frame. Each panel also shows yellow Agent suggestion tags on the right and a Confirm button.}
\end{figure}

The Seed Panel is VideoMaker’s core interaction interface and is used in four key stages: Co-create the Script, Co-create the Storyboard Script, Define Visual Style and Characters, and Generate Images and Videos (Figure~\ref{fig:4}). It externalizes the current creative intent as a visible and editable shared state, while making the collaboration boundary explicit: creator-confirmed seeds define what should remain, while AI explores what can still vary. The interface consists of three areas: Seed Space, AI Agent, and Result Preview.\par

\begin{itemize}
\item Seed Space externalizes and manages key information that creators want to preserve. Depending on the task, seeds may include script keywords, storyboard keywords, style reference images, or style descriptors. Together, these seeds constitute the current shared and editable state of creative intent.

\item AI Agent reads the current task state, interacts continuously with the creator, and proactively proposes candidate seeds for aspects that remain unspecified.

\item Result Preview displays scripts, storyboards, images, or other outputs generated from the current seed configuration, allowing creators to inspect how the current intent is reflected in generated results and revise seeds accordingly.

\end{itemize}

Using \textit{Before the Silence} as an example, after the creator enters the initial intent, the AI Agent first externalizes explicit information into initial seeds such as “young composer,” “mother has died,” “discovers old cassette tapes,” and “fragmented recordings,” turning the initial idea into an explicit intent state that can be inspected and edited. It then retrieves possible story development structures from the Multi-style Script Library and generates an initial script based on these seeds.\par

Creators can directly renegotiate the current generation boundaries by editing seeds. For example, if “mother has died” constrains the story too early, the creator can revise it to “mother has been missing for years.” The AI Agent then restructures the relevant plot around the updated seeds. This illustrates that creator-confirmed intent can be revised as the creative process evolves. Creators can also ask the AI Agent for suggestions through dialogue. In addition, when the system detects prolonged interaction without further specification, the AI Agent proactively proposes new ideas as candidate seeds. Only candidate seeds confirmed by the creator become part of the current intent state, forming a continuous negotiation cycle.\par

\subsection{AI Video Authoring Workflow for Progressive Intent Translation}
\label{sec:4.5}

VideoMaker organizes creation from initial intent to final video into six sequential stages: Input Creative Idea \ensuremath{\rightarrow} Co-create the Script \ensuremath{\rightarrow} Co-create the Storyboard Script \ensuremath{\rightarrow} Define Visual Style and Characters \ensuremath{\rightarrow} Generate Images and Videos \ensuremath{\rightarrow} Editing and Integration (Figure~\ref{fig:4}). This staged structure progressively translates creative intent from an abstract idea into increasingly concrete narrative, audiovisual, and generative representations. Information is carried across stages through the shared project context, AI Agent, and Seed Panel, supporting the progressive translation of creative intent throughout the workflow.\par

\section{Evaluation}
\label{sec:5}

This section presents our evaluation of VideoMaker. We recruited 30 participants with varying levels of creative experience. The evaluation focused on VideoMaker’s support for differentiated expansion of initial intent (DG1), progressive and faithful translation of creative intent across stages and representations (DG2), and continuously negotiable boundaries between creator-confirmed intent and AI initiative (DG3). We examined whether the system could support the expansion, translation, and negotiation of creative intent throughout human–AI video co-creation.\par

\subsection{Participants}
\label{sec:5.1}

We recruited 30 eligible participants. All were at least 18 years old, consented to audio recording throughout the study, and reported no visual or auditory impairments that would interfere with the video creation tasks. Participants were anonymized as P01–P30. Seventeen were male and 13 were female, aged 18–26 years (\textit{M} = 20.73). All participants had prior experience with generative AI tools, with 18 reporting almost daily use. Their AI video creation experience ranged from no prior use to frequent use in real-world projects. Detailed information is shown in Table~\ref{tab:5}.\par

\begin{table}[!htbp]
\centering
\caption{Participant demographics and prior experience with generative AI and AI video creation.}
\label{tab:5}
\Description{Demographics table for 30 evaluation participants (P01-P30). Columns: ID, gender, age, generative AI use frequency, AI video experience. Summary: 17 male, 13 female; ages 18-26 (mean 20.73); 18 participants use generative AI almost daily. Row data: P01 F age 26 AI daily video tried 0 works; P02 M age 22 AI weekly video 4+ works; P03 M age 24 AI daily video tried 0 works; P04 F age 21 AI weekly video never; P05 M age 26 AI daily video 4+ works; P06 F age 19 AI daily video tried 0 works; P07 M age 18 AI monthly video 1-3 works; P08 M age 24 AI daily video never; P09 M age 19 AI 1-5 times video never; P10 F age 20 AI daily video tried 0 works; P11 M age 20 AI weekly video tried 0 works; P12 F age 23 AI daily video never; P13 M age 21 AI weekly video 4+ works; P14 M age 20 AI daily video 1-3 works; P15 F age 20 AI daily video tried 0 works; P16 F age 21 AI weekly video 1-3 works; P17 M age 20 AI daily video 1-3 works; P18 F age 20 AI daily video 1-3 works; P19 M age 19 AI daily video tried 0 works; P20 M age 20 AI daily video 1-3 works; P21 M age 20 AI daily video 1-3 works; P22 M age 20 AI daily video 4+ works; P23 M age 20 AI weekly video 4+ works; P24 M age 20 AI daily video 4+ works; P25 F age 19 AI monthly video 1-3 works; P26 F age 21 AI weekly video never; P27 F age 19 AI weekly video 4+ works; P28 F age 20 AI monthly video 1-3 works; P29 F age 20 AI daily video 1-3 works; P30 M age 20 AI daily video 1-3 works;}
\begingroup
\scriptsize
\setlength{\tabcolsep}{3pt}
\renewcommand{\arraystretch}{1.08}
\begin{tabularx}{\linewidth}{@{}>{\raggedright\arraybackslash\hsize=0.6747\hsize}X>{\raggedright\arraybackslash\hsize=0.8672\hsize}X>{\raggedright\arraybackslash\hsize=0.5583\hsize}X>{\raggedright\arraybackslash\hsize=1.2503\hsize}X>{\raggedright\arraybackslash\hsize=1.6496\hsize}X@{}}
\toprule
ID & Gender & Age & Generative AI Use & AI Video Experience \\
\midrule
P01 & Female & 26 & Almost daily & Tried a few times (0 work) \\
P02 & Male & 22 & Weekly & Created 4+ works \\
P03 & Male & 24 & Almost daily & Tried a few times (0 work) \\
P04 & Female & 21 & Weekly & Never used \\
P05 & Male & 26 & Almost daily & Created 4+ works \\
P06 & Female & 19 & Almost daily & Tried a few times (0 work) \\
P07 & Male & 18 & Monthly & Created 1–3 works \\
P08 & Male & 24 & Almost daily & Never used \\
P09 & Male & 19 & Used 1–5 times & Never used \\
P10 & Female & 20 & Almost daily & Tried a few times (0 work) \\
P11 & Male & 20 & Weekly & Tried a few times (0 work) \\
P12 & Female & 23 & Almost daily & Never used \\
P13 & Male & 21 & Weekly & Created 4+ works \\
P14 & Male & 20 & Almost daily & Created 1–3 works \\
P15 & Female & 20 & Almost daily & Tried a few times (0 work) \\
P16 & Female & 21 & Weekly & Created 1–3 works \\
P17 & Male & 20 & Almost daily & Created 1–3 works \\
P18 & Female & 20 & Almost daily & Created 1–3 works \\
P19 & Male & 19 & Almost daily & Tried a few times (0 work) \\
P20 & Male & 20 & Almost daily & Created 1–3 works \\
P21 & Male & 20 & Almost daily & Created 1–3 works \\
P22 & Male & 20 & Almost daily & Created 4+ works \\
P23 & Male & 20 & Weekly & Created 4+ works \\
P24 & Male & 20 & Almost daily & Created 4+ works \\
P25 & Female & 19 & Monthly & Created 1–3 works \\
P26 & Female & 21 & Weekly & Never used \\
P27 & Female & 19 & Weekly & Created 4+ works \\
P28 & Female & 20 & Monthly & Created 1–3 works \\
P29 & Female & 20 & Almost daily & Created 1–3 works \\
P30 & Male & 20 & Almost daily & Created 1–3 works \\
\bottomrule
\end{tabularx}
\endgroup
\end{table}

\subsection{Procedure}
\label{sec:5.2}

We adopted a within-subject design to reduce the influence of individual differences in creative experience and AI familiarity~\cite{li2026creatorcentric,lee2025mvprompt}. All 30 participants completed one short-video creation task with both VideoMaker and the Baseline system, with system order randomized across participants.\par

The two systems supported different workflows. VideoMaker organizes AI-assisted video creation into a structured process centered on the Multi-style Script Library, Cinematic Language Knowledge Base, Seed Panel, and cross-stage AI Agent. For comparison, we built a Baseline system that approximated common AI video creation workflows. Within a single workspace, it provided three basic functions: text-based dialogue, image generation, and video generation. It did not provide knowledge augmentation, explicit intent externalization through the Seed Panel, or cross-stage state propagation. Rather than using a real-world toolchain composed of multiple third-party products, we constructed this Baseline to control for confounds such as tool switching, interface differences, and prior familiarity, thereby focusing the comparison on VideoMaker’s core mechanisms \cite{li2026creatorcentric}.\par

Before the study, the researcher explained the study purpose, task requirements, informed-consent procedures, and the think-aloud protocol. At the beginning of each condition, participants randomly selected a creation topic and were asked to produce a playable AI-generated short video of at least 15 seconds within 30 minutes. After each condition, participants completed a post-condition questionnaire (Appendix~\ref{sec:C}). After completing both conditions, they participated in a semi-structured interview about their overall experience, system preferences, and related issues (Appendix~\ref{sec:D}).\par

Following the main study, to examine whether the above process-level support translated into viable video outputs, three experts in video creation (one female, two male; average age 26.7; distinct from the main study participants) conducted a quality assessment of all the narrative videos produced with VideoMaker. Before rating each video, they were provided with its initial concept and creator-confirmed intent, and then evaluated the final output with brief rationales.\par

\subsection{Measures and Analysis}
\label{sec:5.3}

We analyzed data from the post-condition questionnaires, think-aloud sessions, and semi-structured interviews. Both conditions used the same 12 seven-point Likert-scale items (Appendix~\ref{sec:C}). C.1 included the six CSI dimensions — Exploration, Collaboration, Engagement, Effort/Reward Tradeoff, Tool Transparency, and Expressiveness \cite{carroll2009csi} — adapted to the context of AI-assisted video creation, with the response scale adapted to a 7-point Likert scale. In C.2, Exploration and Differentiation corresponded to DG1; Narrative Translation and Intent Expressiveness to DG2; Intent Preservation and Negotiable AI Initiative to DG3. The questionnaire also included six comparative items (C.3) measuring system preference and willingness to use.\par

For the quantitative analysis, we treated the 30 participants' responses as paired observations. All ratings were valid integers from 1 to 7 with no missing data. For each matched item, we report the mean and standard deviation (Mean \ensuremath{\pm} SD) for both conditions. Differences between conditions were evaluated using two-sided paired Wilcoxon signed-rank tests, with Holm correction applied across the 12 primary tests. The core statistical distributions, error bars, and significance levels are visually summarized in the figures (Figures~\ref{fig:5}--\ref{fig:7}). The six comparative items (C.3) were tested against the scale midpoint of 4 with Holm correction, using two-sided one-sample Wilcoxon signed-rank tests, interpreted as comparative preference and willingness to use.\par

For the qualitative analysis, recordings from the semi-structured interviews and think-aloud sessions were transcribed verbatim, checked against the original audio, de-identified, and labeled by participant ID, system condition, and timestamp. Two researchers then conducted a hybrid thematic analysis using a codebook that combined the deductive DG1–DG3 framework with data-driven codes. They first independently coded approximately 20\% of the material, reconciled disagreements, and refined the codebook. They then coded the full dataset, grouped related codes into broader categories, and synthesized them into higher-level themes, while noting divergent perspectives.\par

Finally, we integrated quantitative results, qualitative themes, and representative quotations through a DG1–DG3 evidence mapping process. This analysis examined convergence, complementarity, and tension across evidence sources, triangulating questionnaire ratings with interview feedback and linking each finding to specific system components and design goals.\par

The expert assessment used two 7-point rating items assessing the following questions. Expert ratings were summarized descriptively and interpreted together with their rationales.\par

\begin{itemize}
\item Q1: To what extent did the final video accurately realize the creator's confirmed core creative intent and narrative structure following the cross-stage translation?

\item Q2: To what extent did the short video demonstrate a complete narrative framework and basic audiovisual presentation at the level of an initial rough cut?

\end{itemize}

\section{Findings}
\label{sec:6}

This study draws on quantitative analysis of questionnaire data and qualitative coding of interviews. The findings are organized into three main parts: (1) support for the design goals; (2) creative support and interaction experience; and (3) expectations for future systems.\par

Overall, VideoMaker demonstrated significant advantages in the differentiated expansion of initial creative intent and its narrative translation across stages, several dimensions of the creative experience and willingness to use, while providing significantly stronger support for preserving creator-confirmed intent. At the same time, the negotiated boundaries between creator-confirmed intent and AI initiative could not yet be maintained consistently. Future systems should go beyond making creative intent “visible and editable” to ensure that it can also be continuously carried forward, continuously negotiated, and reliably enforced.\par

\subsection{Support for the Design Goals}
\label{sec:6.1}

The design of VideoMaker is grounded in three design goals derived from the formative study, respectively addressing intent expansion (DG1), intent translation (DG2), and  creator-AI negotiation (DG3). These three design goals address how creative intent can be expanded, translated, and negotiated as an evolving shared state throughout AI-assisted video creation.\par

\begin{figure}[!htbp]
\centering
\includegraphics[width=\linewidth]{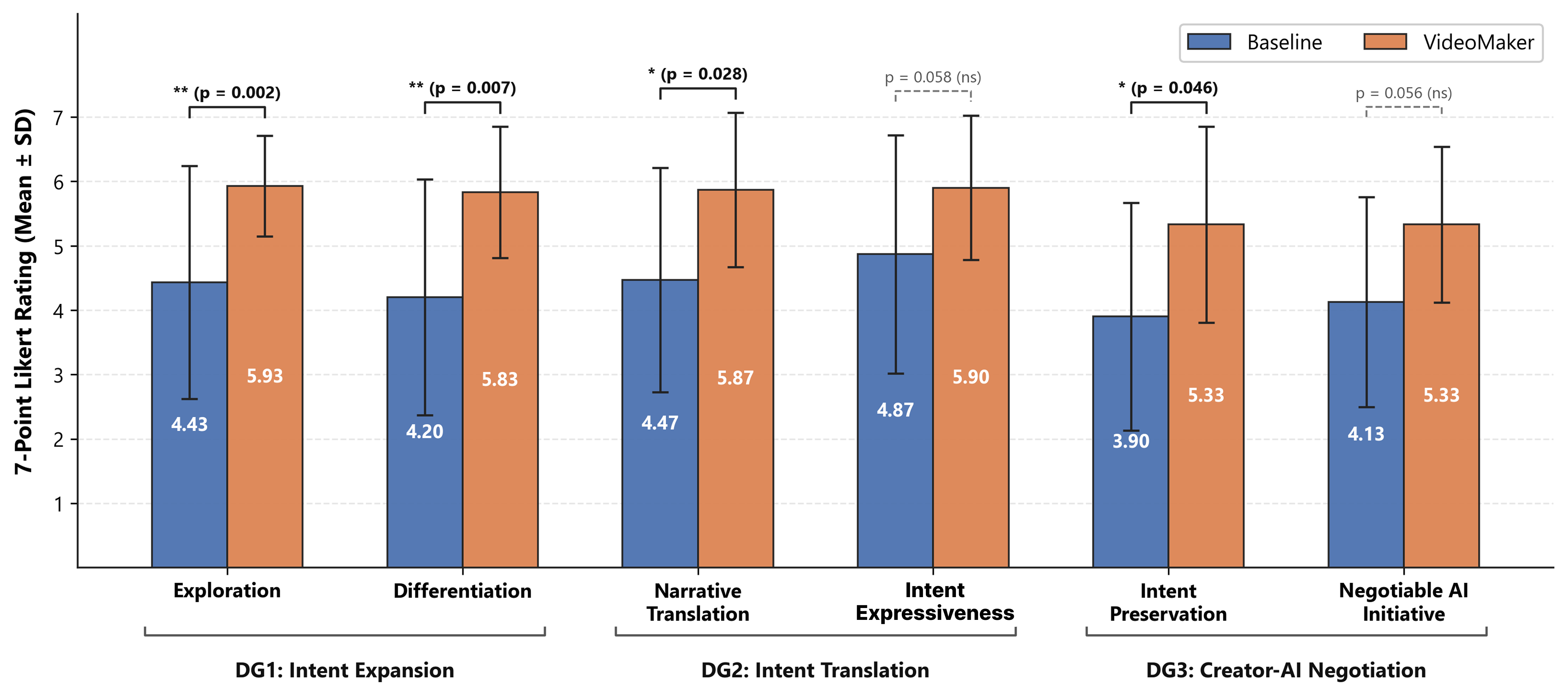}
\caption{Ratings of design goals support, comparing VideoMaker with the Baseline system.}
\label{fig:5}
\Description{Grouped bar chart titled design-goal support ratings, comparing Baseline and VideoMaker. The vertical axis shows 7-point Likert ratings (mean plus or minus standard deviation), from 0 to 7. Six items are grouped under DG1 Intent Expansion (Exploration, Differentiation), DG2 Intent Translation (Narrative Translation, Intent Expressiveness), and DG3 Creator-AI Negotiation (Intent Preservation, Negotiable AI Initiative). VideoMaker is higher on all six items; four comparisons are statistically significant after Holm correction. Exploration: Baseline 4.43, VideoMaker 5.93, p equals .002. Differentiation: Baseline 4.20, VideoMaker 5.83, p equals .007. Narrative Translation: Baseline 4.47, VideoMaker 5.87, p equals .028. Intent Expressiveness: Baseline 4.87, VideoMaker 5.90, p equals .058, not significant. Intent Preservation: Baseline 3.90, VideoMaker 5.33, p equals .046. Negotiable AI Initiative: Baseline 4.13, VideoMaker 5.33, p equals .056, not significant.}
\end{figure}

\subsubsection{Intent Expansion Support (DG1)}
\label{sec:6.1.1}

Compared with the Baseline system's single prompt-based input, VideoMaker leverages the Multi-style Script Library and the Seed Panel to expand an abstract initial intent into multiple discrete and visible candidate branches during the early creative stage.\par

Questionnaire results indicated that participants perceived VideoMaker as significantly better at Exploration and Differentiation (Exploration: M = 5.93 vs. 4.43, p = .002; Differentiation: M = 5.83 vs. 4.20, p = .007; see Figure~\ref{fig:5} DG1). Twenty-two participants noted that the Seed Panel provided unexpected creative directions and enabled them to compare, discard, and recombine ideas, thereby helping them break free from their initial thinking and discover creative possibilities they had not actively considered. For example, P16 mentioned that the initial directions provided by the AI served as “a set of benchmarks,” upon which they could further “diverge,” effectively broadening their creative thinking. These findings suggest that the system supported differentiated expansion of the initial creative intent.\par

\subsubsection{Intent Translation Support (DG2)}
\label{sec:6.1.2}

The Baseline system lacked inspectable, structured intermediate states. In contrast, VideoMaker established a standardized AI Video Authoring Workflow and incorporated the Cinematic Language Knowledge Base, with an AI Agent running throughout the workflow that progressively translates natural-language ideas into structured narratives and professional audiovisual representations that carry across stages. Questionnaire results indicated improved translation support: participants rated VideoMaker significantly higher than the Baseline system on Narrative Translation (M = 5.87 vs. 4.47, p = .028; see Figure~\ref{fig:5} DG2). Eighteen participants reported that the staged workflow helped them understand story development and translate vague ideas into concrete shot descriptions. P07 noted that the system-generated descriptions were “highly professional in terminology and vividly imagistic,” suggesting that structured representations lowered the barrier to professional expression.\par

However, the Intent Expressiveness item did not reach the corrected significance threshold (M = 5.90 vs. 4.87, p = .058; see Figure~\ref{fig:5} DG2). This was primarily because structured decomposition, while improving the clarity of individual shots, introduced semantic fragmentation. Several participants reported that the atmosphere across consecutive shots lacked coherence — for instance, P08 described their expression as having been “chopped into discrete pieces.” These findings suggest that progressive translation improved the explicitness of local representations, while maintaining faithful continuity across stages and shots remains challenging.\par

\subsubsection{Creator-AI Negotiation Support (DG3)}
\label{sec:6.1.3}

In the Baseline system, creators relied mainly on repeatedly rewriting full-length prompts to adjust the generation space, making it difficult to explicitly distinguish what should remain preserved from what remained open to exploration. VideoMaker instead uses the Seed Panel to externalize creator-confirmed information as editable seeds, while allowing the AI Agent to proactively propose candidate seeds in unspecified areas. This enables creators to continuously adjust the boundary between preserved intent and AI initiative. Questionnaire results showed that VideoMaker was rated significantly higher on Intent Preservation (M = 5.33 vs. 3.90, p = .046; see Figure~\ref{fig:5} DG3). For Negotiable AI Initiative, VideoMaker showed a positive directional advantage, but the difference did not reach the corrected significance threshold (M = 5.33 vs. 4.13, p = .056; see Figure~\ref{fig:5} DG3). Seventeen participants (17/30) affirmed the role of discrete seeds in maintaining a sense of creative direction. They described how anchoring core creative intent in confirmed seed labels allowed them to preserve the narrative thread without stifling the AI's associative capacity. P18 characterized this interaction as “a coupling of intent anchoring and divergent generation”: even with key constraints in place, the system could still deliver “creation across more dimensions.”\par

However, Negotiable AI Initiative did not show a consistently reliable advantage. Some participants noted that previously excluded features occasionally reappeared in later shots, or that localized adjustments unexpectedly perturbed adjacent frames. When AI's divergence exceeded expectations, the negotiated boundary established by the creator was not reliably maintained. P30 stated bluntly that in certain stages the system's “creativity outweighs controllability,” causing the final video to deviate from their original vision. These findings suggest that DG3 is not about establishing a one-time balance between human control and AI freedom. Rather, the boundary between what should persist and what remains open to AI initiative must remain continuously adjustable as creation stages evolve.\par

\subsection{Creative Experience, Workflow Practices, and Output Quality}
\label{sec:6.2}

\begin{figure}[!htbp]
\centering
\includegraphics[width=0.94\linewidth]{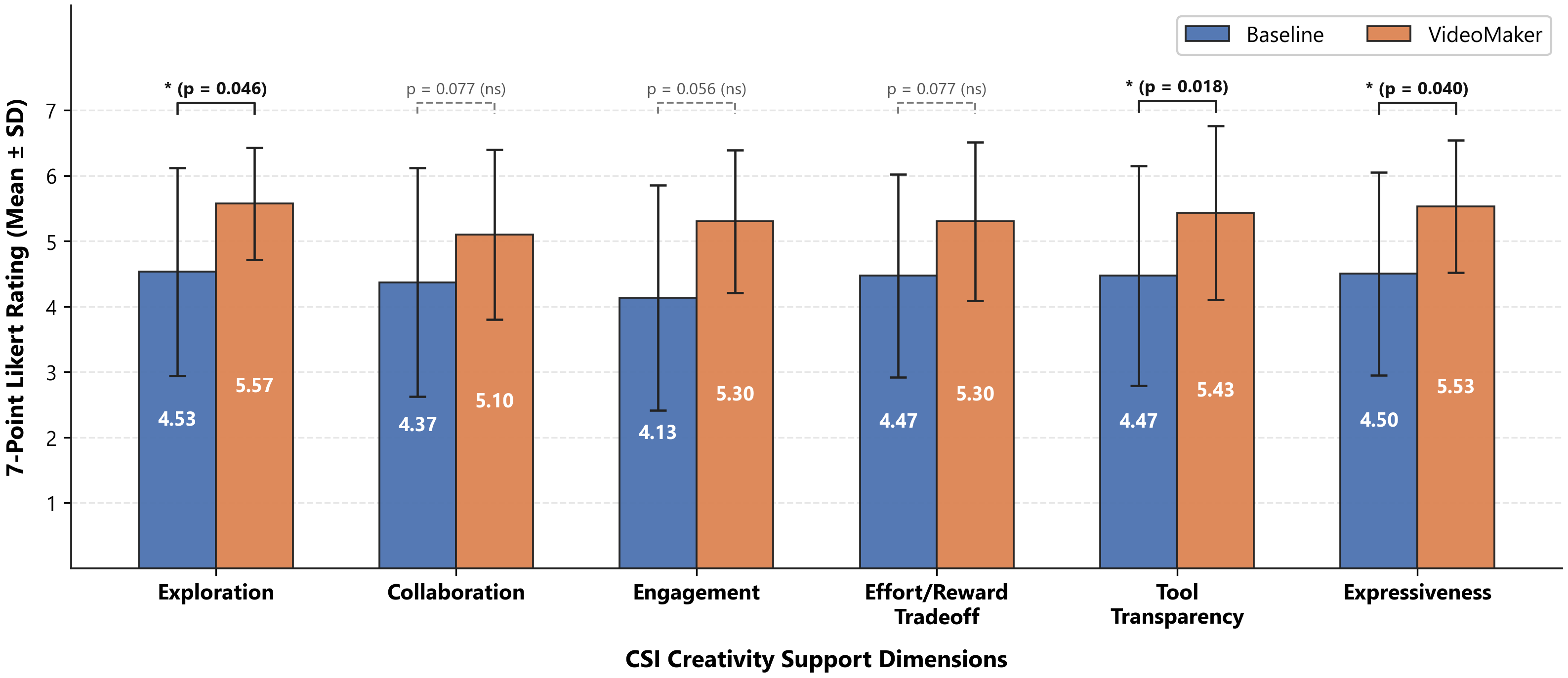}
\caption{Ratings of creativity support dimensions (CSI), comparing VideoMaker with the Baseline system.}
\label{fig:6}
\Description{Grouped bar chart of Creativity Support Index ratings, comparing Baseline and VideoMaker. The vertical axis shows 7-point Likert ratings (mean plus or minus standard deviation), from 1 to 7. Six dimensions are shown: Exploration, Collaboration, Engagement, Effort/Reward Tradeoff, Tool Transparency, and Expressiveness. VideoMaker is higher on all six dimensions; three are statistically significant. Exploration: Baseline 4.53, VideoMaker 5.57, p equals .046. Collaboration: Baseline 4.37, VideoMaker 5.10, p equals .077, not significant. Engagement: Baseline 4.13, VideoMaker 5.30, p equals .056, not significant. Effort/Reward Tradeoff: Baseline 4.47, VideoMaker 5.30, p equals .077, not significant. Tool Transparency: Baseline 4.47, VideoMaker 5.43, p equals .018. Expressiveness: Baseline 4.50, VideoMaker 5.53, p equals .040.}
\end{figure}

\begin{figure}[!htbp]
\centering
\includegraphics[width=0.88\linewidth]{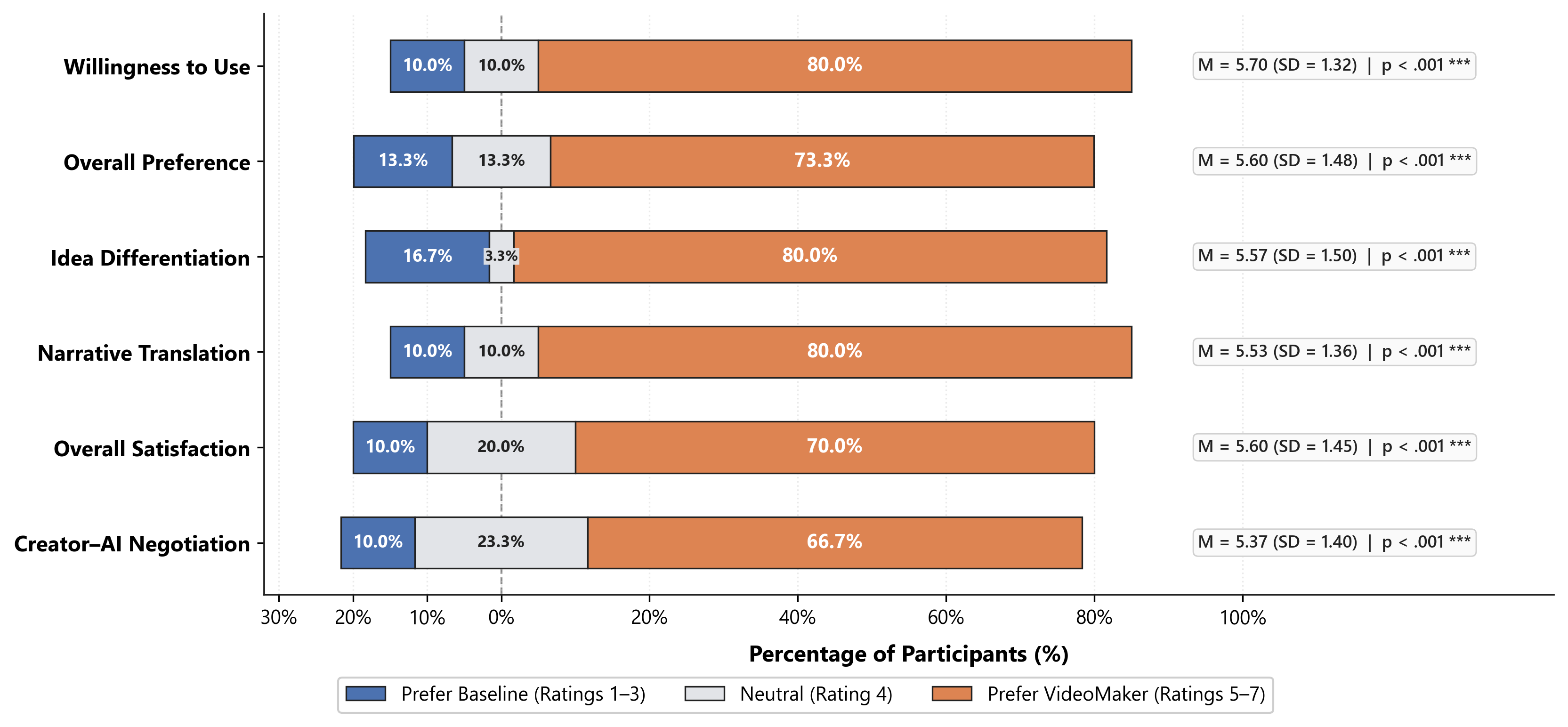}
\caption{Ratings of overall satisfaction and preference, comparing VideoMaker with the Baseline system.}
\label{fig:7}
\Description{Diverging stacked bar chart of comparative preference ratings for 30 participants. Six items appear on the vertical axis: Willingness to Use, Overall Preference, Idea Differentiation, Narrative Translation, Overall Satisfaction, and Creator-AI Negotiation. Each horizontal bar shows the percentage preferring Baseline (ratings 1-3), neutral (rating 4), or VideoMaker (ratings 5-7). All six items are significantly above the midpoint of 4 (all p less than .001). A majority preferred VideoMaker on every item (66.7 to 80.0 percent). Willingness to Use: 10.0 percent Baseline, 10.0 percent neutral, 80.0 percent VideoMaker; mean 5.70, SD 1.32. Overall Preference: 13.3, 13.3, 73.3; mean 5.60, SD 1.48. Idea Differentiation: 16.7, 3.3, 80.0; mean 5.57, SD 1.50. Narrative Translation: 10.0, 10.0, 80.0; mean 5.53, SD 1.36. Overall Satisfaction: 10.0, 20.0, 70.0; mean 5.60, SD 1.45. Creator-AI Negotiation: 10.0, 23.3, 66.7; mean 5.37, SD 1.40.}
\end{figure}

To evaluate the two systems' overall performance beyond the core design goals, we further examined differences in creativity support dimensions (CSI, see Figure~\ref{fig:6}) and overall satisfaction and preference (see Figure~\ref{fig:7}), while drawing on interviews to examine process visibility, traceability, the cognitive overhead introduced by the staged workflow and workflow trade-offs under time-bounded creation.\par

\subsubsection{Significantly Stronger Exploration, Tool Transparency, and Expressiveness}
\label{sec:6.2.1}

In the creativity support evaluation (Appendix~\ref{sec:C.1}), VideoMaker scored significantly higher than the Baseline system on Exploration (M = 5.57 vs. 4.53, p = .046), Tool Transparency (M = 5.43 vs. 4.47, p = .018), and Expressiveness (M = 5.53 vs. 4.50, p = .040) (see Figure~\ref{fig:6}). Positive directional advantages were also observed for Collaboration (M = 5.10 vs. 4.37, p = .077), Effort/Reward Tradeoff (M = 5.30 vs. 4.47, p = .077), and Engagement (M = 5.30 vs. 4.13, p = .056). The continuous process feedback effectively reduced the frustration of blind trial and error and encouraged creators to engage actively with concrete audiovisual decisions.\par

The interviews further illuminated these experiential differences. Seventeen participants pointed out that VideoMaker made the creative trajectory coherent and visible, allowing them to intuitively track how each adjustment influenced subsequent frames. P30 summarized the stage-by-stage progression as “every part was deeply immersive,” emphasizing that they could clearly perceive how their ideas took shape step by step, which freed up more attention for refining audiovisual details. This feedback also reflects VideoMaker’s design of externalizing creative intent and allowing it to evolve visibly across stages.\par

That said, the staged workflow also introduced cognitive overhead. Thirteen participants mentioned that the multiple modules and specialized terminology required an adaptation period. P16 remarked candidly that the initial operation was “slightly complex.” This suggests that the gains from creative support are moderated by onboarding cost, and that simplifying the functional hierarchy is key to lowering the entry barrier.\par

\subsubsection{A Visible Creative Process Supports Higher Satisfaction and Preference}
\label{sec:6.2.2}

Results from the relative preference scale (C.3) further demonstrated VideoMaker’s overall advantage (see Figure~\ref{fig:7}): all dimensions were significantly above the midpoint of 4 (all p < .001). Willingness to Use (M = 5.70, 80.0\% > 4) and Overall Preference (M = 5.60, 73.3\% > 4) were particularly pronounced, with more than 70\% of participants expressing a clear preference for VideoMaker in both long-term use intention and overall experience.\par

Seventeen participants attributed their high satisfaction to the visibility and traceability of the overall creative process. P18 noted that the system made visible “the process of my thinking about making a video,” so that every attempt left a traceable record. P23 added that the staged progression clarified their thinking — “each step only requires focusing on the current content” — which effectively reduced compound cognitive interference.\par

Preferences also had clear boundaries. Twelve participants noted that VideoMaker was better suited to multi-shot, complex narratives, while the Baseline system felt more lightweight for simple ideas (P20: “for longer-form stories it would definitely be better”). This suggests that treating creative intent as a shared state that evolves across stages may be particularly valuable for complex tasks, while the structured workflow can introduce unnecessary overhead for simpler tasks.\par

\subsubsection{Output Quality Under Time-Constrained Creation}
\label{sec:6.2.3}

\begin{figure}[!htbp]
\centering
\includegraphics[width=0.88\linewidth]{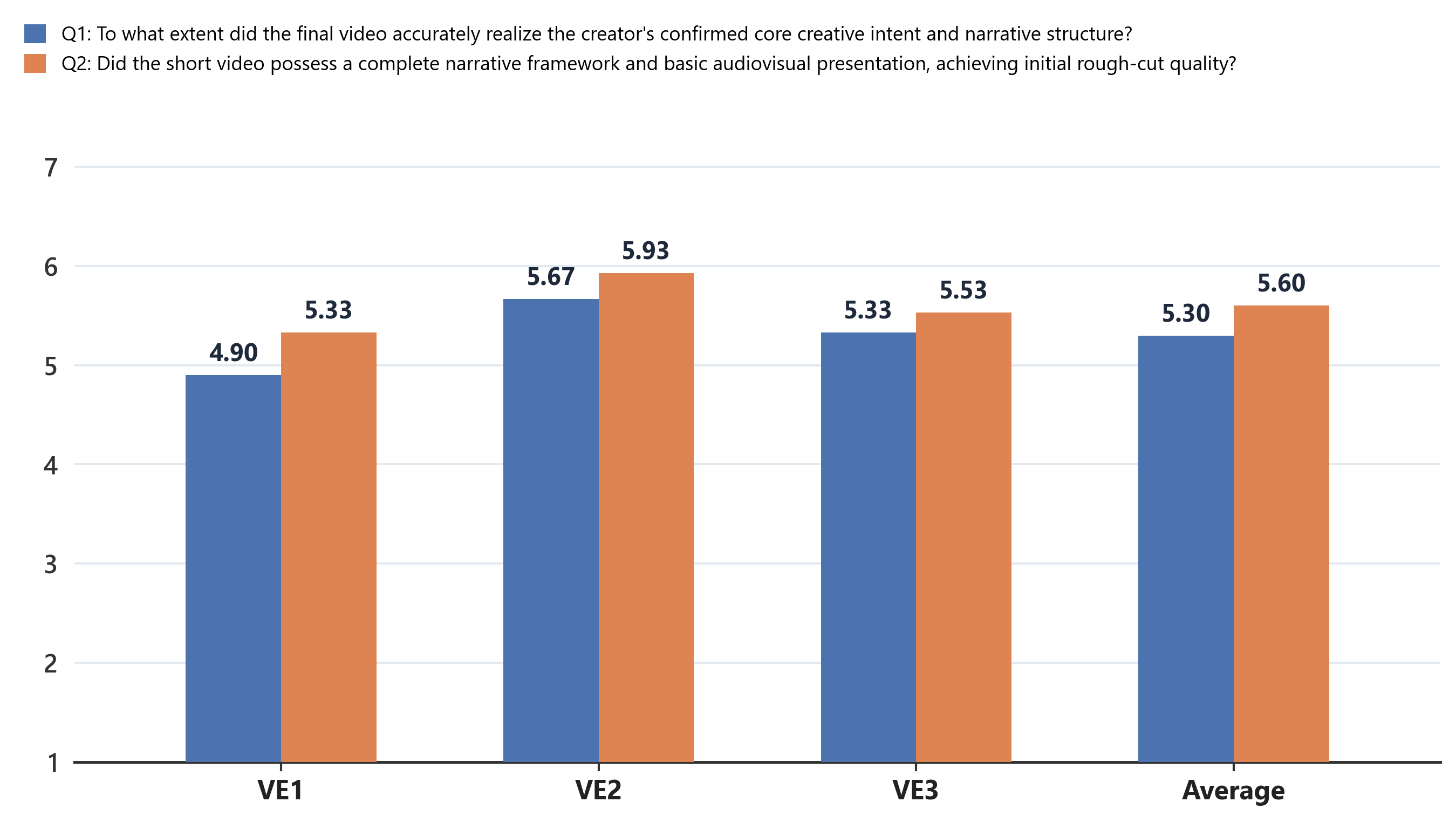}
\caption{Average ratings provided by experts for the intent realization and rough-cut completeness of the narrative shorts.}
\label{fig:8}
\Description{Grouped bar chart of expert ratings for intent realization and rough-cut completeness. The vertical axis shows 7-point ratings. Four groups appear on the horizontal axis: VE1, VE2, VE3, and Average. Q1 measures intent realization; Q2 measures rough-cut completeness. Q2 is higher than Q1 in every group. VE2 has the highest scores; VE1 the lowest. VE1: Q1 4.90, Q2 5.33. VE2: Q1 5.67, Q2 5.93. VE3: Q1 5.33, Q2 5.53. Average: Q1 5.30, Q2 5.60.}
\end{figure}

\begin{figure}[!htbp]
\centering
\includegraphics[width=\linewidth]{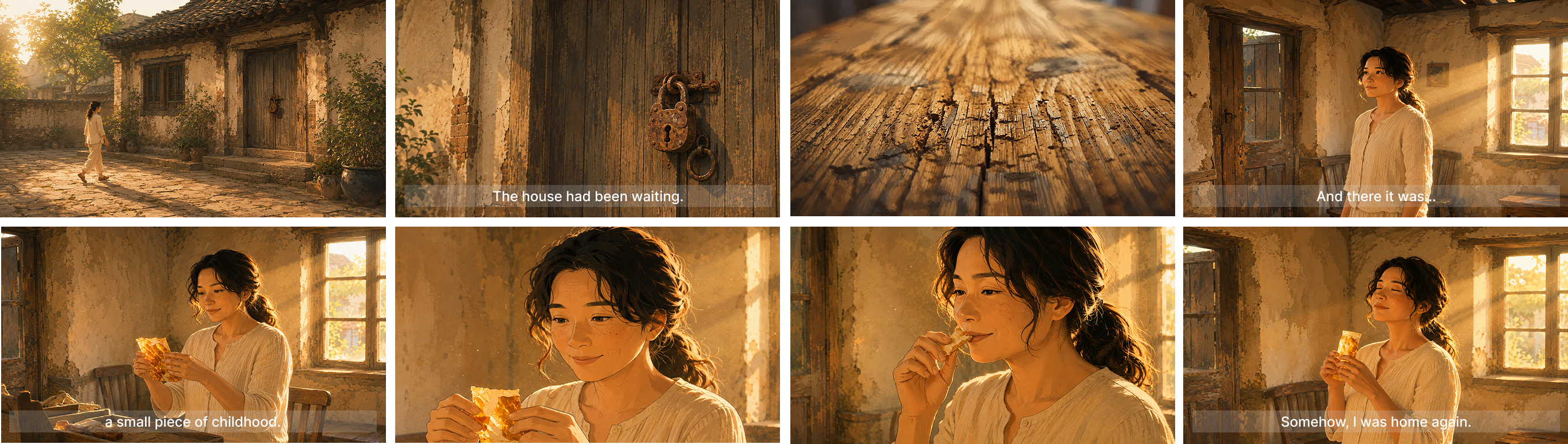}
\caption{Representative frames from \textit{The House That Remembered}, a narrative short created with VideoMaker during the study.}
\label{fig:9}
\Description{Eight-frame storyboard sequence in two rows of four panels. Panels follow chronological order with warm golden-hour lighting and show a woman returning to an old home. Panel 1: she walks across a sunlit stone courtyard toward a rustic house. Panel 2: close-up of a weathered wooden door with a rusty padlock; on-screen text reads, The house had been waiting. Panel 3: extreme close-up of cracked aged wood grain. Panel 4: she stands in a dusty interior as sunlight streams through a window; on-screen text reads, And there it was. Panel 5: she sits at a table holding a small golden snack wrapper; on-screen text reads, a small piece of childhood. Panel 6: close-up of her face looking down at the food. Panel 7: close-up as she takes a bite with eyes closed. Panel 8: she sits back with eyes closed and a peaceful smile; on-screen text reads, Somehow, I was home again.}
\end{figure}

The expert ratings (see Figure~\ref{fig:8}) showed average scores of approximately 5.3 and 5.6 for the two metrics, respectively. These findings indicate that, within a 30-minute time constraint, the VideoMaker-supported six-stage workflow enabled creators to produce structurally complete and playable initial narrative shorts (see Figure~\ref{fig:9}), though the quality remained largely at the level of a “usable rough cut.” Due to the limited creation time, there was still room for improvement regarding the precision of individual shot details, cross-shot consistency, and the nuanced conveyance of creative atmosphere—further highlighting the practical gap between rapid prototyping and in-depth refinement.\par

\subsection{Participant Expectations for Future Systems}
\label{sec:6.3}

Drawing on interview feedback collected during the system evaluation, we identified three directions for deeper improvement. These reflect limitations and opportunities for expansion in the current prototype regarding cross-stage global consistency, precisely scoped control, and direct multimodal and professional editing capabilities.\par

\subsubsection{Cross-Stage and Cross-Shot Global Consistency}
\label{sec:6.3.1}

In the system evaluation, half of the participants (15/30) identified global consistency as the most pressing need. Although seeds could be set up front, generation stochasticity still frequently caused visual features and narrative tone to drift. The interviews revealed multidimensional breakdown points. At a deeper level, this stemmed from a sense of “semantic slicing” introduced by stage-wise decomposition: although individual-shot audiovisual descriptions became more professional, the shots lacked cross-temporal contextual anchoring, making lighting, character appearance, and emotional tone prone to discontinuity. For instance, P26 wanted stronger cross-shot coherence to prevent character appearance from “changing dramatically”; P30 noted that the final video was “discontinuous from shot to shot”; and P11 reported that background style varied too greatly across scenes, undermining the unity of the story space. To maintain consistency, participants had to re-enter lengthy prompts for every shot, imposing a heavy operational burden.\par

To address this bottleneck, participants suggested establishing project-level global rules that fix characters, scenes, and styles as cross-shot shared constraints, with automatic synchronization of modifications and visual highlighting of affected scope. This indicates that the system needs stronger global state management across the timeline to safeguard long-range audiovisual narrative coherence.\par

\subsubsection{Reversible, Intent-Preserving, and Well-Scoped Control}
\label{sec:6.3.2}

Eight participants (8/30) emphasized the need for reversibility and isolation of local operations. Currently, the tight coupling of multi-stage states means that fine-tuning a single shot can sometimes unexpectedly affect other shots or overwrite earlier seeds. P13 described this uncontrolled coupling as frustrating: “after changing shot B, shot A also changed on its own.” P03 expressed a desire for localized refinement, wishing to narrow the scope of modifications and have “memory to roll back.” P04 likewise noted that the absence of version history made trial-and-error too risky. Participants broadly wanted local-modification isolation barriers to ensure that adjustments remain strictly confined within a single shot.\par

In addition, participants suggested dynamically modulating AI initiative by creation stage: offering divergent suggestions during ideation, then switching to an “intent-preserving mode” once the direction is set. This suggests that effective co-creative control depends on predictable, reversible, and well-scoped coordination mechanisms that encourage experimentation while protecting creator autonomy.\par

\subsubsection{Direct Multimodal and Professional Editing Capabilities}
\label{sec:6.3.3}

Six participants (6/30) wanted the workflow to extend into direct multimodal refinement and professional editing. The current prototype relies on text-triggered regeneration, making it difficult to convey fine-grained audiovisual editing intent precisely. For visual editing, participants expected visual reference and localized inpainting mechanisms based on keyframes. For instance, P02 noted a desire to make local modifications based on an existing satisfactory image, rather than regenerating the entire frame each time. For audio, P06 suggested that beyond voiceover narration, the system should support direct character dialogue, while P30 mentioned that adding ambient sound effects would better convey the on-site atmosphere. Furthermore, participants expressed a need for multi-track timeline editing to directly drag assets, fine-tune audio–video alignment, and layer multiple audio tracks.\par

These findings suggest that creative intent should remain editable not only during ideation and generation, but also during final refinement and post-production. By integrating reference-image input, local brushes, and multi-track timelines, the system could substantially shorten the distance from creative intent to finished output, enabling creative intent to be continuously refined from generation to final production.\par

\section{Discussion}
\label{sec:7}

\subsection{Design Implications for Human-AI Video Co-creation Systems}
\label{sec:7.1}

\subsubsection{Transforming AI-Generated Outputs into an Explorable and Actionable Creative Space}
\label{sec:7.1.1}

Although generative AI can provide a large number of candidate outputs, creators still need to compare, select, and combine different creative directions \cite{derooij2026homogeneous,anderson2024homogenization,suh2024luminate,fan2024contextcam}. This study found that the Seed Panel in VideoMaker transforms generated outputs into comparable creative units~\cite{zhou2024nonlinear}, allowing creators to explore new possibilities around their existing creative intent rather than passively accepting generated outcomes.\par

Participants did not treat AI-generated content as final solutions. Instead, they used these outputs as references for further creation. Interview results showed that participants particularly valued AI’s ability to offer unexpected creative directions while retaining the ability to independently select and modify them. For example, some participants reported that the Seed Panel helped them discover story elements they had not considered, yet they still needed to reorganize and refine these elements according to personal preferences. This suggests that the value of AI-generated outputs lies not only in higher generation success rates, but also in constructing a creative space where creators can explore possibilities. Future AI video systems could further support candidate comparison, exploration trajectory preservation, element anchoring, and version branching, enabling creators to organize AI-generated outputs in a manner similar to managing creative sketches.\par

\subsubsection{Transforming Cinematic Knowledge into Progressive Creative Scaffolding}
\label{sec:7.1.2}

Related studies have shown that structured representations and domain knowledge can support users in complex creative tasks \cite{antony2024id8,wang2024reelframer,wen2025seewhat}. This study further demonstrates that cinematic knowledge can serve as an intermediate scaffold between creative intent and professional audiovisual expression. Interviews further revealed that participants were not lacking creative intent, but often struggled to translate their ideas into cinematic language. For example, participants could describe abstract goals such as “expressing loneliness” or “creating a cinematic feeling,” yet found it difficult to determine camera choices, pacing, and visual organization. The script structure, storyboard fields, and cinematic terminology explanations in VideoMaker provided an intermediate bridge between abstract concepts and concrete expressions. Importantly, the value of cinematic knowledge support extends beyond reducing immediate creative barriers. It may also help users gradually develop an understanding of camera language and narrative structures.\par

Therefore, future AI video systems should serve not only as generation tools but also as progressive creative scaffolding environments, helping creators accumulate expressive abilities through practice. For beginners, systems can provide explanations of camera language, reference examples, and structural suggestions. For experienced creators, systems can reduce auxiliary information and offer greater editing flexibility.\par

\subsubsection{Designing AI Initiative as a Stage-Adaptive Negotiation Process}
\label{sec:7.1.3}

Our findings suggest that the challenge is not only to establish negotiable boundaries between creator-confirmed intent and AI initiative, but also to adapt these boundaries to the changing needs of different creative stages~\cite{zhang2025agencyreview,kadenhe2025codesign}. In particular, when AI divergence exceeded participants’ expectations, the negotiated boundaries established earlier in the process could not always be reliably maintained. As P30 noted, in some stages “creativity outweighs controllability,” causing the final video to deviate from their intended vision.\par

This points to an opportunity for future AI video systems to make AI initiative stage-adaptive. Rather than maintaining a fixed configuration of the Seed Panel throughout the workflow, the system could adjust which intents remain persistent and which aspects remain open to AI exploration according to the current creative stage. For example, during early creative exploration, more content could remain open to AI initiative, whereas during later script refinement, storyboard adjustment, and generation, confirmed seeds could be given stronger persistence to reduce unintended divergence. Such stage-adaptive negotiation could provide a more flexible way to retain creator-led decision making while preserving AI’s capacity for creative exploration.\par

\subsubsection{Making Evolving Creative Intent Visible, Traceable, and Editable}
\label{sec:7.1.4}

Generative AI systems often emphasize final output quality \cite{ma2025videodiffusion}, while this study found that creators also care about how creative intent is interpreted, translated, and revised throughout generation. When videos were generated directly, participants often experienced uncertainty about “why this result was produced,” whereas structured intermediate representations  and seeds helped them understand how AI transformed abstract ideas into concrete audiovisual expressions. This suggests that constructing creative intent as a shared state requires more than having the system “remember” it; the state and its evolution must also remain visible, traceable, and editable to creators. Future AI video systems should therefore emphasize process visibility and make key intermediate states available as editable objects. For example, systems could visualize the evolution of different creative directions, record transformations from stories to shots, and allow creators to return to previous stages for revision. This design could transform AI from an opaque generator into a collaborative creative environment in which creators can actively participate.\par

However, expanding the creative space also introduces selection costs. Some participants reported that too many candidate directions could cause thematic drift or difficulty in prioritizing options. This suggests that creative exploration should not simply increase the number of generated outputs, but should also provide mechanisms for direction filtering, version comparison, and intention preservation.\par

\subsection{Rethinking AI-assisted Video Creation: From Generation Tools to Creative Process Infrastructure}
\label{sec:7.2}

AI-assisted video creation should not be understood as a one-time process of “text input–video generation,” but rather as a continuous human-AI co-creation process.\par

First, the value of AI video systems is shifting from improving generation efficiency toward supporting the entire creative process. Previous studies suggest that the contribution of AI creative tools depends on their position within creative workflows and whether they support continuous creator participation in creative decisions \cite{hwang2022toolate,kadenhe2025codesign}. Therefore, future system design should focus on how creators explore, understand, modify, and control AI-generated results. Second, the processes of intent expansion, translation, and negotiation identified in this study suggest that AI video creation is an iterative process between divergence and convergence. When evaluating human-AI co-creative systems, researchers should consider not only final output quality, but also whether creators can understand AI suggestions, accept or reject generated results, and continuously develop and calibrate creative intent across multiple interactions \cite{zhang2025agencyreview,carroll2009csi,haase2024cocreativity}. Finally, the value of AI video systems is influenced by task complexity and creator experience. For simple tasks, automatic generation may already satisfy user needs. However, for long-form videos, multi-shot narratives, and complex storytelling tasks, creators require AI support for structural organization, knowledge assistance, and process management \cite{antony2024id8,wang2024reelframer}. Future evaluation frameworks should therefore incorporate process-oriented metrics, including interaction trajectories, version changes, and intention preservation, rather than relying solely on final artifacts.\par

\subsection{Limitations and Future Work}
\label{sec:7.3}

While this study provides insights into AI-assisted video creation, further investigation is needed to examine how these findings generalize across diverse users and creative contexts \cite{lakier2024knickknacks,gong2026doubleedged}. Participants in this study were primarily recruited from university populations, resulting in relatively concentrated age and educational backgrounds. In addition, experimental tasks mainly focused on short-video creation scenarios. Future work could involve creators with diverse levels of experience and explore VideoMaker in more authentic and long-term creative practices.\par

This study primarily evaluated participants’ perceived experiences through self-reported measures. Although CSI can capture users’ perceived creative support, it remains a subjective evaluation approach and should be complemented with objective behavioral evidence \cite{carroll2009csi}. Future research could incorporate system logs, interaction trajectories, and generated version histories to further examine how creative intent evolves across stages and how creators accept, modify, or reject AI suggestions during the creative process.\par

This study compared only the full VideoMaker system with the Baseline system and did not conduct ablation studies on core features of VideoMaker. Although measures such as the CSI can capture system performance across different experiential dimensions, they cannot further distinguish the independent contributions of individual features. The observed improvements in individual dimensions may therefore reflect the combined effects of multiple features, and attributing these improvements to any single feature may overestimate its individual contribution.\par

This study examined short-term human-AI co-creation experiences within a controlled experimental setting. Future research could further investigate how long-term interaction with AI systems influences creators’ workflows, creative habits, and collaboration patterns. Potential directions include cross-project memory, personalized creative preference modeling, and finer-grained adjustment of AI initiative.\par

Overall, VideoMaker demonstrates a possible pathway from generation tools toward creative collaboration systems. Future AI video systems should not merely improve automatic generation capabilities, but should continuously support creators in understanding, controlling, and developing their own creative processes.\par

\section{Conclusion}
\label{sec:8}

This study investigates how creative intent can be constructed as an evolving shared state in cross-stage creator–AI collaboration for video creation. Interviews with 12 creators informed 3 design goals: supporting differentiated expansion around creative intent; supporting the progressive and faithful translation of creative intent across stages and representations; and supporting continuously negotiable boundaries between creator-confirmed intent and AI initiative. We then co-designed and developed VideoMaker with three experts. The system organizes the process as a six-stage progressive translation workflow, using the Multi-style Script Library to support differentiated expansion of initial intent, the Cinematic Language Knowledge Base to support faithful translation, and the Seed Panel to externalize creator-confirmed intent as a shared state, making what should remain preserved and what should remain open to AI exploration continuously negotiable. Overall, this study suggests that future AI video systems should not merely pursue stronger generative capabilities, but should further support the continuous evolution of creative intent throughout cross-stage co-creation.\par

\section{Acknowledgments of the Use of AI}
\label{sec:9}

We utilized artificial intelligence, specifically large language models and multimodal generative models, to support both system implementation and the research process. The VideoMaker prototype integrates a large language model (Qwen3.6-35B-A3B) alongside image, video, and speech generation models to perform the following: (1) script ideation and structured storyboard generation across the six-stage workflow; (2) automated integration of cross-stage generation prompts and seed extraction; and (3) multimodal asset synthesis, including storyboard images, video clips, and voiceover narration. For transparency, the system prompts used at each stage are provided in Appendix~\ref{sec:B}. During the evaluation study, participants interacted with these AI-powered features, and the results reflect their authentic perceptions of AI-assisted video creation. The authors take full responsibility for all uses of artificial intelligence in this work.\par

\section*{Acknowledgments}

We sincerely thank the video creators who participated in our formative interviews and the 30 participants in our user evaluation study for their valuable time and insightful feedback, which laid the foundation for the design and assessment of VideoMaker.\par

\appendix

\section{Semi-Structured Interview Protocol of Formative Study}
\label{sec:A}

Follow-up questions were asked as appropriate. Background information was collected before the formal interview.\par

\subsection{Participant Background and AI Video Creation Experience}
\label{sec:A.1}

\begin{itemize}
\item Please introduce your age, occupation, and educational background, and whether you work in content creation or video production.

\item Have you created short-form videos or animations before? Please describe a recent or representative experience.

\item How would you describe your experience with existing AI short-video tools?

\end{itemize}

\subsection{Existing AI Video Creation Practices and Workflows}
\label{sec:A.2}

\begin{itemize}
\item When you begin with only a vague topic or idea, how do you usually develop it further?

\item Do you use AI for ideation? If so, how?

\item What intermediate content do you usually create between an initial idea and the actual video generation?

\item When creation spans multiple stages, how do you determine whether later content still aligns with your initial idea?

\item Have you experienced cases where the initial idea was clear but the final output did not align with it? Why?

\item When AI-generated results do not meet your expectations, how do you usually revise them?

\end{itemize}

\subsection{Needs and Challenges in a Situated Ideation Task}
\label{sec:A.3}

\begin{itemize}
\item Please conceive an AI-generated animated short video based on the given topic and describe your thought process.

\item During this process, which steps would you complete yourself, and where would you seek AI assistance?

\item Which parts of the process do you find most difficult or effort-intensive?

\end{itemize}

\subsection{Expectations for AI Creative Assistants and Collaboration}
\label{sec:A.4}

\begin{itemize}
\item If AI could participate throughout the entire process, where would you most want its assistance?

\item Which content or decisions would you want to determine yourself, and which could be delegated directly to AI?

\item How would you like to accept, reject, revise, combine, or revert AI-generated suggestions?

\item What would your ideal mode of collaboration with AI in video creation look like?

\end{itemize}

\section{Prompts for VideoMaker’s Six Stages}
\label{sec:B}

Note: The original prompts are lengthy and are condensed here for presentation.\par

\subsection{Input Creative Idea}
\label{sec:B.1}

\begin{itemize}
\item Convert the user’s idea into seeds: Decompose the user’s input into key information such as characters, scenes, plot, emotion, and visual features; recommend a script style from the Multi-style Script Library; and output the information as categorized seeds.

\end{itemize}

\subsection{Co-create the Script}
\label{sec:B.2}

\begin{itemize}
\item Explain professional terms: Refer to the Cinematic Language Knowledge Base to briefly explain the selected script seed or professional term, including its definition, key techniques, and role in the script.

\item Generate a story from seeds: Create a complete short story based on the creator-confirmed script seeds and script-style seed, preserving all confirmed elements and reflecting the selected script style.

\item Generate a textual storyboard script: Refer to the definition and structure of textual storyboard scripts in the Cinematic Language Knowledge Base, decompose the story into scenes and shots, and generate a detailed visual description and voiceover for each shot.

\end{itemize}

\subsection{Co-create the Storyboard Script}
\label{sec:B.3}

\begin{itemize}
\item Generate a structured storyboard: Adapt a specified shot from the textual storyboard script into an animation storyboard entry, including shot number, visual content, voiceover, duration, depth of field, shot size, camera angle, and camera movement.

\item Generate an image prompt from storyboard fields: Integrate the current shot’s visual content, camera angle, depth of field, shot size, character, and style information into a complete Chinese image-generation prompt.

\item Extract seeds from an individual storyboard shot: Extract key information including visuals, audio, duration, depth of field, shot size, camera angle, and camera movement, and output them as categorized storyboard seeds.

\item Explain cinematic terms in the storyboard: Refer to the Cinematic Language Knowledge Base to explain the selected visual or cinematic-language term, including its definition and function in the current shot.

\item Integrate an image prompt from storyboard seeds: Generate a complete Chinese image-generation prompt from the creator-confirmed storyboard seeds, ensuring that all confirmed visual elements are included.

\end{itemize}

\subsection{Define Visual Style and Characters}
\label{sec:B.4}

\begin{itemize}
\item Identify key characters: Identify the key characters from the complete storyboard, rank them by narrative importance, and limit the number to three.

\item Generate character seeds: Describe each key character’s identity, worldview, personality, expression, facial features, hairstyle, body type, and distinctive traits, and organize them into categorized character seeds.

\item Generate a character image prompt: Translate abstract character seeds into concrete appearance descriptions and integrate them into a complete prompt for generating a character reference image.

\item Recommend visual styles: Based on the story, recommend four matching visual styles from the system’s visual style knowledge base and rank them by relevance.

\end{itemize}

\subsection{Generate Images and Videos}
\label{sec:B.5}

\begin{itemize}
\item Generate storyboard image prompts: Combine the storyboard description, cinematic parameters, storyboard seeds, confirmed character seeds, and confirmed style seeds into the final image-generation prompt.

\item Identify characters in the shot: Based on the current storyboard description, determine which confirmed key characters appear in the shot. If none appear, output “None.”

\item Extract video dynamics: Extract information related to actions, motion, and change from the storyboard description and camera movement, and integrate it into a video dynamics description.

\item Extract video dynamics seeds: Extract dynamic seeds including camera movement, subject motion, environmental motion, speed and rhythm, emotional pauses, and action transitions.

\item Generate a video prompt: Generate a complete video-generation prompt from the creator-confirmed dynamics seeds while preserving all confirmed motion information.

\end{itemize}

\subsection{Editing and Integration}
\label{sec:B.6}

\begin{itemize}
\item Generate music suggestions: Based on the completed storyboard and generated video content, recommend background music in terms of emotion, style, genre, and rhythmic characteristics.

\item Plan shot transitions: Based on the complete storyboard and the system’s transition library, determine which adjacent shots require transitions and select an appropriate transition type for each.

\end{itemize}

\section{Post-Condition Questionnaire}
\label{sec:C}

Both conditions included the same 12 items rated on a 7-point Likert scale (six items in C.1 and six in C.2), where 1 = Strongly Disagree and 7 = Strongly Agree. The VideoMaker condition additionally included six comparative items against the Baseline system (C.3).\par

\subsection{Creativity Support Index (CSI)}
\label{sec:C.1}

\begin{itemize}
\item Exploration: It was easy for me to explore many different options, ideas, designs, or outcomes without a lot of tedious, repetitive interaction.

\item Collaboration: I was able to work together with others easily while doing this activity.

\item Engagement: I was very absorbed/engaged in this activity — I enjoyed it and would do it again.

\item Effort/Reward Tradeoff: What I was able to produce was worth the effort required to produce it.

\item Tool Transparency: While I was doing the activity, the tool/interface/system “disappeared,” and I was able to concentrate on the activity.

\item Expressiveness: I was able to be very expressive and creative while doing the activity.

\end{itemize}

\subsection{Design-Goal Experience Evaluation}
\label{sec:C.2}

\begin{itemize}
\item Exploration: The system supports me in exploring diverse ideas, options, and directions for video creation.

\item Differentiation: The system supports me in developing differentiated video concepts.

\item Narrative Translation: The system supports the cross-stage translation of natural-language ideas into structured video narratives, including scenes, shots, and their sequence.

\item Intent Expressiveness: The system allows me to clearly express my ideas without relying on professional filmmaking terminology.

\item Intent Preservation: The system allows me to clearly specify and preserve the parts of my ideas that should remain unchanged.

\item Negotiable AI Initiative: The system allows me to adjust what AI may proactively explore.

\end{itemize}

\subsection{Overall Comparison and Intention to Use}
\label{sec:C.3}

\begin{itemize}
\item Overall Preference: Compared with the Baseline system, I would prefer to use this system for future AI video projects.

\item Compared with the Baseline system, this system performs better in supporting differentiated and diverse ideas.

\item Compared with the Baseline system, this system performs better in translating natural-language ideas into video narratives.

\item Compared with the Baseline system, this system performs better in supporting negotiation between me and AI.

\item I would be willing to use this system in real-world video creation projects.

\item Overall, I am more satisfied with this system than with the Baseline system.

\end{itemize}

\section{Evaluation Semi-Structured Interview Protocol}
\label{sec:D}

\begin{itemize}
\item Overall, what were the biggest differences between the two creation experiences? Please describe each system in two or three words.

\item For questionnaire items showing notable differences between conditions, participants were asked to explain the reasons for their ratings.

\item Which system would you prefer to use in a real project?\newline{}\textit{A subset of the following questions was asked as appropriate:}

\item Which system made it easier to explore different creative directions? Can you give an example of when you actually changed an idea or style?

\item How did you turn the ideas in your mind into scenes and shot sequences?

\item At which stages did you feel that you could control the pace and direction of creation?

\item In which creation experience did you feel more engaged and more willing to continue?

\item Which system was easier to learn? Which required more mental effort or caused more frustration? Please give a specific example for each.

\item Suggestions

\end{itemize}

\textit{A subset of the following questions was asked as appropriate:}\par

\begin{itemize}
\item What improvements would you suggest for the system?

\item How would you evaluate your experience using the Seed Panel?

\item How would you like AI-generated content to be edited or refined after generation?

\item Is there anything else you would like to add?

\end{itemize}
\bibliographystyle{ACM-Reference-Format}
\bibliography{references}
\end{document}